\documentclass[fleqn,usenatbib]{mnras}

\usepackage{newtxtext,newtxmath}

\usepackage[T1]{fontenc}
\usepackage{ae,aecompl}

\usepackage{graphicx}	
\usepackage{amsmath}	
\usepackage{amssymb}	
\usepackage{subfigure}
\usepackage{float}
\usepackage{mathrsfs}
\usepackage{caption}

\def\eagle{{\sc eagle}}

\title[Quenching timescales in \eagle\ ]{Quenching timescales of galaxies in the \eagle\ simulations}

\author[R. J. Wright et al.]{Ruby J. Wright$^{1,2}$, Claudia del P. Lagos$^{1,2}$, Luke J. M. Davies$^{1}$, Chris Power$^{1,2}$, \newauthor{James W. Trayford$^{3}$, O. Ivy Wong$^{1,2}$}
\\
$^{1}$International Centre for Radio Astronomy Research (ICRAR), M468, University of Western Australia, 35 Stirling Hwy, Crawley, \\WA 6009, Australia\\
$^{2}$ARC Centre of Excellence for All Sky Astrophysics in 3 Dimensions (ASTRO 3D).\\
$^{3}$Leiden Observatory, Leiden University, PO Box 9513, NL-2300 RA Leiden, the Netherlands.}

\date{Accepted XXX. Received YYY; in original form ZZZ}

\pubyear{2019}

\begin{document}
\label{firstpage}
\pagerange{\pageref{firstpage}--\pageref{lastpage}}
\maketitle

\begin{abstract}
We use the \eagle\ simulations to study the connection between the quenching timescale, $\tau_{\rm Q}$, and the physical mechanisms that transform star-forming galaxies into passive galaxies. By quantifying $\tau_{\rm Q}$ in two complementary ways - as the time over which (i) galaxies traverse the green valley on the colour-mass diagram, or (ii) leave the main sequence of star formation and subsequently arrive on the passive cloud in specific star formation rate (SSFR)-mass space - we find that the $\tau_{\rm Q}$ distribution of high-mass centrals, low-mass centrals and satellites are divergent. In the low stellar mass regime where $M_{\star}<10^{9.6}M_{\odot}$, centrals exhibit systematically longer quenching timescales than satellites ($\approx 4$~Gyr compared to $\approx 2$~Gyr). Satellites with low stellar mass relative to their halo mass cause this disparity, with ram pressure stripping quenching these galaxies rapidly. Low mass centrals are quenched as a result of stellar feedback, associated with long $\tau_{\rm Q}\gtrsim 3$~Gyr. At intermediate stellar masses where $10^{9.7}\,\rm M_{\odot}<M_{\star}<10^{10.3}\,\rm M_{\odot}$, $\tau_{\rm Q}$ are the longest for both centrals and satellites, particularly for galaxies with higher gas fractions. At $M_{\star}\gtrsim 10^{10.3}\,\rm M_{\odot}$, galaxy merger counts and black hole activity increase steeply for all galaxies. Quenching timescales for centrals and satellites decrease with stellar mass in this regime to $\tau_{\rm Q}\lesssim2$~Gyr. In anticipation of new intermediate redshift observational galaxy surveys, we analyse the passive and star-forming fractions of galaxies across redshift, and find that the $\tau_{\rm Q}$ peak at intermediate stellar masses is responsible for a peak (inflection point) in the fraction of green valley central (satellite) galaxies at $z\approx 0.5-0.7$.
\end{abstract}





\section{Introduction}

Large observational galaxy surveys have demonstrated a distinct bimodality in the galaxy colour-magnitude diagram (CMD) (e.g. \citealt{Strateva2001,Baldry2006,WYDER07,TAYLOR15}). At low redshift, two distinct populations of galaxies emerge separated by their colour, accordingly denoted the red sequence and the blue cloud. The red sequence is tight, and typically home to  early-type galaxies with little star formation activity. The blue cloud is comparatively more dispersed, and is normally home to late-type galaxies with higher star formation rates \citep{Baldry2004,Driver2006,Schiminovich07}. Between these two populations is a region denoted the green valley, which appears in observations to be sparsely populated \citep{Martin2007, Salim2014} (though work by \citet{Eales2018} demonstrates that observational biases could be responsible for the under-density). Galaxy bimodality can also be observed in SSFR-stellar mass space \citep{WYDER07}, where the blue cloud population in colour-mass space is correlated with the main sequence population in the SSFR-stellar mass plane \citep{Noeske07,Davies2016GAMASFR}. 

Since the green valley is nominally found to be sparsely populated, it has been theorised that galaxies evolve from the blue cloud to the red sequence rapidly as their star formation rates slow, the colour transformation being indicative of a process broadly referred to as star formation quenching  (e.g. \citealt{Blanton2006,Borch2006,Bundy2006}, \citealt{Fang2013,Moustakas2013,Salim2014,Peng2015,Bremer18}). Star formation quenching can be the result of one or a combination of mechanisms which can be classified as internal or external. An internal quenching mechanism stifles star formation due to some galaxy-localised process, while an external quenching mechanism slows star formation in a galaxy by some environmentally-driven effect. Internal quenching mechanisms include, but are not limited to: active galactic nuclei (AGN) activity \citep{DiMatteo2005,Murray05,Croton2006,Bower06,Hopkins2006,Lagos08},
supernovae feedback \citep{Springel2005a,Cox2006a,DallaVecchia2012,Lagos13}, and
virial shock heating \citep{Birnboim2003,Keres2009}. External quenching mechanisms include, but are not limited to: major \& minor mergers \citep{Toomre1972,Springel2005b,Cox2006b,McNamara2006}, ram pressure stripping \citep{Gunn1972,Knobel2013,Brown2017}, and galaxy strangulation \citep{Larson1980,Balogh2000,Knobel2013,Peng2015}. Although we make a clear distinction between these quenching mechanisms, the reader should note that quenching mechanisms can be, and often are, interrelated in a non-linear manner \citep{Hopkins2008p}. Most notable is the relationship between AGN activity and galaxy merging events, where angular momentum dissipation acts to funnel matter towards a galaxy's central super-massive black hole (SMBH) and up-regulate black hole accretion rates  \citep{Barnes1991,Mihos1996,Hopkins2008a,Kocevski2012}. 

Each quenching mechanism, external or internal, could act over different, distinctive timescales. Any quenching mechanism acts by depleting a galaxy's cool ISM gas reservoir otherwise available for star formation. Since mechanisms of gas removal or heating could act over different timescales, the quenching mechanism will determine how long it takes the galaxy to quench. \citep{Dave12,Lilly13}. This idea of a gas reservoir in galaxies acting to regulate star formation is commonly referred to as the ``bathtub'' model of gas supply in galaxies, which explains how the quantity and properties of gas in a galaxy can change over time through various inflows and outflows. 

The primary observational method used to constrain quenching timescales of galaxies (${\tau}_{\textrm{Q}}$) -  the time it takes for a galaxy to transform from star-forming to passive - draws on an exponentially decaying model of star formation rate (SFR) \citep{Wetzel2013,Schawinski2014,Hahn2017,Smethurst2018}. \citet{Wetzel2013} found a characteristic e-folding time for star formation in satellite galaxies of ${\tau}_{\rm Q}<0.8$~Gyr, $2-4$~Gyr after infall to a group environment. \citet{Schawinski2014} used SDSS DR7 data to show that there are two distinct pathways that galaxies take in quenching and traversing the green valley, characterised by different quenching timescales. Early-type galaxies were found to have shorter e-fold quenching timescales (${\tau}_{\textrm{Q}}<250$ Myr), while late-type galaxies would quench more slowly with timescales ${\tau}_{\textrm{Q}}\approx 1$~Gyr. \citet{Hahn2017} use SDSS DR7 data to calculate the quenching timescales of galaxies with stellar masses  $10^{9.5}M_{\odot}<M_{\star}<10^{11}M_{\odot}$. They found characteristic e-folding times of $0.5-1.5$~Gyr, implying a total migration time from the star formation main sequence to quiescence on the order of $\approx4$~Gyr, with central galaxies typically taking $2$~Gyr longer to quench than satellites. \citet{Smethurst2018} also used SDSS DR7 data to investigate the quenching timescales of slow- and fast-rotators, finding that slow-rotators would quench rapidly (with $\tau_{\rm Q}<1$~Gyr) compared to fast-rotators (which displayed more spread in $\tau_{\rm Q}$). Using arguments based on the density of green-valley galaxies in the GAMA survey, \citet{Bremer18} found a relatively universal crossing timescale for galaxies with stellar masses  $10^{10.25}M_{\odot}<M_{\star}<10^{10.75}M_{\odot}$, at $z<0.2$ of ${\tau}_{\rm Q}\approx 1-2$~Gyr. This is in contrast to some of the aforementioned results, which show varying quenching timescales based on galaxy properties. The recent observational work of \citet{HerreraCamus2019} used a different method of constraining quenching timescales in observations. They make physically motivated arguments regarding energy dissipation and gas reservoir depletion timescales, together with spatially resolved data of a $z=2$ galaxy (zC400528) to show that its AGN-driven outflow is capable of expelling central molecular gas in an ``inside-out'' manner over a timescale of $\approx 0.2$~Gyr.

Moving away from observational studies, multiple large scale projects involving the hydrodynamical simulation of galaxy formation have found great success in reproducing several observations of galaxy populations, in recent years becoming state-of-the-art in their capabilities as predictive tools for extra-galactic astrophysics. Such simulations include \eagle\ \citep{SCHAYE2015,CRAIN2015,McAlpine2016}, Illustris-TNG \citep{Pillepich18}, Horizon-AGN \citep{Dubois16}, MUFASA \citep{Dave16}, and others.

Some tension exists in the results arising from these new simulations concerning the color bimodality and quenching timescales of galaxies. 
\citet{TRAYFORD2016} used \eagle\ to study the colour evolution of galaxies, concluding that galaxies join the red sequence because either (1) they become satellites in a larger system (the path affecting mostly lower mass galaxies), or (2) because of AGN feedback (the path affecting mostly higher mass galaxies). Using the Horizon-AGN simulation, \citet{Dubois16} show that star formation quenching, galaxy morphology, and cosmic gas accretion are intricately related, where early-type galaxies require AGN feedback to down-regulate cosmic gas infall and cease star-formation to regrow disk structures (see also \citealt{Sparre17}). These different mechanisms act on different stellar mass ranges, and should naturally lead to different quenching timescales. 

In apparent contrast to these conclusions, \citet{Nelson2018} found in the Illustris-TNG simulations that the colour transition timescale of galaxies (using the colour-mass diagram) was unimodal, on average $\approx 1.6$~Gyr over all stellar mass bins from $M_{\star}=10^{9}\,\rm M_{\odot}$ to $10^{12.5}\,\rm M_{\odot}$, with higher mass galaxies typically showing shorter quenching timescales. The unimodal distribution of quenching timescales they observed opposes the existence of multiple colour transition pathways with distinct physical timescales. This said, their work does demonstrate a small peak in quenching timescales in for $M_{\star}\approx 10^{10}M_{\odot}$ for the TNG100-1 run for both centrals and satellites. They also found that in general, the behaviour of centrals and satellites in their colour transition were very similar. 

The aim of this work is to study the quenching timescales of galaxies produced by the \eagle\ suite of cosmological simulations using multiple methods, and to connect the distribution of ${\tau}_{\rm Q}$ to the physical mechanisms which stifle star formation. In order to maximise the robustness of our approach, we employ two very different methods of quantifying ${\tau}_{\rm Q}$: (i) by analysing the movement of galaxies through colour-mass space (e.g. \citealt{TRAYFORD2016,Nelson2018}), and (ii) making use of the movement of galaxies in the SSFR-stellar mass plane. The diverse galaxy population available in \eagle\ makes this simulation (and numerical models of cosmological galaxy formation in general) an ideal test-bed for identifying and quantifying how certain physical mechanisms would produce changes in the distribution of quenching timescale. In doing so, we add to the developing body of knowledge looking to explain how galaxy colour bimodality in the local Universe has developed, from a mechanistic perspective.

The content we present in this paper is structured as follows. In $\S$~\ref{eaglesec} we introduce the \eagle\ simulation suite and sub-grid models that are relevant to this study, as well as how specific physical quantities were extracted from the database and post-processed. $\S$~\ref{quenchingsec} isolates the effect of external and internal mechanisms on quenching timescale distributions. In $\S$~\ref{conclusions} we conclude what these findings tell us about the physical mechanisms acting to quench different galaxies. 

\section{The \eagle\ Simulations}\label{eaglesec}
Studying quenching timescales using a simulation based approach is ideal as it allows for tracking of individual galactic properties over time, without the need for sweeping assumptions regarding galaxy star formation histories which are otherwise necessary to generate observational calculations of ${\tau}_{\rm Q}$. The \eagle\ (Evolution and Assembly of GaLaxies and their Environments) simulation suite \citep{SCHAYE2015,CRAIN2015} is a collection of cosmological hydrodynamical simulations which follow the evolution of galaxies and cosmological structure from $z=20$ to $z=0$. In \eagle, dark matter (DM) halos are identified using a field-of-friends (FOF) approach \citep{Davis1985}, and subsequently the SUBFIND algorithm \citep{Springel2001,Dolag2009} identifies overdensities of particles within these structures, corresponding to galaxies. The ANARCHY \citep{SCHALLER2015} set of refinements were implemented on the GADGET-3 tree-SPH (smoothed particle hydrodynamics) code \citep{Springel2005a} to perform the \eagle\ simulations over a variety of periodic volumes, at a number of numerical resolutions. \eagle\ adopts the parameters of a ${\Lambda}$CDM universe from \citet{Planck2014}, with initial conditions outlined in \citet{Jenkins2013}. Twenty-nine discrete snapshots of pertinent simulation and post-processed galaxy integrated quantities were taken in each simulation run. For our study, we make use of the Ref-L0100N1504 \eagle\ simulation run, which has a co-moving box size of 100 cMpc, the largest of the suite \citep{SCHAYE2015}. This run has $1504^3$ DM particles, and the same initial number of gas particles. For more details see \citet{SCHAYE2015}.

Sub-grid physics modules were implemented to treat the physics which are important for galaxy formation and evolution but that happen below the resolution of the simulation. These include: (i) radiative cooling and photoheating,
(ii) star formation, (iii) stellar evolution and enrichment, (iv) stellar
feedback, and (v) SMBH growth and AGN feedback. In the next section we provide a brief description of how these mechanisms are modelled in \eagle.

\subsection{Subgrid physics}\label{subgridphysics}

In an element-by-element approach, photo-heating and radiative cooling were applied based on the work of \citet{Wiersma2009}. This included the effect of 11 elements which were deemed influential: H, He, C, N, O, Ne, Mg, Si, S, Ca, and Fe \citep{SCHAYE2015}. The effect of radiation from the UV and X-ray background described by \citet{Haardt2001} was implemented on each element individually. 

Star formation occurs almost exclusively in cold interstellar medium (ISM) gas. Since the \eagle\ simulations do not provide the resolution to model cold, interstellar gas on the required scale, a metallicity dependent density threshold is set, above which, star formation was locally permitted \citep{SCHAYE2015}. The simulation also imposes a temperature floor of $8,000$~K, recognising that the cold gas phase was not modelled effectively \citep{SCHAYE2015}. Gas particles are converted to star particles stochastically, with the star formation rate based on tuning a pressure law \citep{Schaye2008} to observations of the relationship between gas and star formation rate surface densities \citep{Kennicutt1998} at $z = 0$. Also considered was the energy feedback from star formation in the form of stellar winds, radiation, and supernovae. This involved a purely thermal energy injection in the form of a temperature boost, ${\Delta}T_{\rm SF} = 10^{6}K$, based on the work of \citet{DallaVecchia2012}, which acts to heat the gas particles sufficiently close to newly formed stars.
 
Often a distinction is made between quasar-mode and radio-mode galaxies host to AGN \citep{Croton2006}, based on their SMBH accretion efficiency. In \eagle, a fixed efficiency was adopted to reduce the number of feedback channels \citep{SCHAYE2015}. This singular efficiency is closest physically to the more active "quasar-mode" AGN, however acting as a maintenance mode in the practice (see \citealt{Bower17} for a discussion). SMBHs are seeded in \eagle\ when a dark matter halo exceeds a virial mass of $10^{10}\,\rm h^{-1} M_{\odot}$, with the seed SMBHs having an initial mass of $10^{5}\,\rm h^{-1} M_{\odot}$. Subsequently, SMBHs can grow via Eddington-limited-accretion \citep{SCHAYE2015}, as well as mergers with other SMBHs, according to work by \citet{Springel2005b}. AGN feedback in \eagle\ involves a purely thermal energy injection into its surroundings. Similar to star formation feedback, this took the form of a purely thermal temperature boost, chosen to be ${\Delta}T_{\rm BH}=10^{8.5}$K \citep{SCHAYE2015}.

\subsection{Integrated quantities}\label{integratedquantities}
At each of the $29$ snapshots, each galaxy is post-processed to produce integrated quantities describing its properties. In this work, we make heavy use of galaxy mass, galaxy colour (specifically $u^{*}-r^{*}$ colour), and star formation rate measurements. We provide a brief overview of how each of these are calculated below. 

We make use of the mass of galaxies calculated in the Aperture, SubHalo and FOF tables of the \eagle\ database. In the Aperture table, mass is calculated by summing the mass of the relevant particles within a spherical aperture centered on the minimum of the gravitational potential of a given galaxy. From the Aperture table we used a $30$~kpc aperture for stellar mass, gas mass (within $30$~kpc) and black hole mass. Star formation rates are also derived from the Aperture table by summing the instantaneous star formation rate intrinsic to gas particles within a $30$~pkpc spherical radius. We also extracted total gas masses of the galaxy using the SubHalo table, which simply sums the mass of all gas particles belonging to a certain SubHalo. Lastly, we also derive halo mass from the FOF table, which sums the mass of all individual particles belonging a certain FOF group. 

Galaxy colour measurements are derived from the Magnitudes table in the database. Magnitudes are calculated in broadband SDSS \citep{Doi2010} filters, with methods described in \citet{Trayford2015}. Briefly, the spectra produced by the GALAXEV population synthesis model \citep{Bruzual2003} are convolved with the broad-band SDSS filters to produce un-obscured broad-band luminosities, which are then used to calculate the broad-band luminosities of full galaxies. These magnitudes are calculated from a $30$~pkpc spherical aperture (which offers the best cutoff between galaxy light and intra-cluster light), and are not corrected for dust attenuation (i.e. are intrinsic) \citep{Trayford2015}. The magnitudes quoted are rest-frame and absolute in the AB system. {\sc eagle} produces a red sequence and blue cloud of galaxies at $z=0.1$ with fractions of galaxies in each that agree well to observations \citep{Trayford2015}. This said, the blue cloud of {\sc eagle} galaxies stretches to higher luminosities than observations, corresponding to the slight underestimate of the passive fraction of massive galaxies in the simulation.

\subsection{Measuring quenching timescales in EAGLE}\label{sec:definitions}

We measure the quenching timescale, $\tau_{\rm Q}$, of galaxies in \eagle\ by tracking their $u^{*}-r^{*}$ ({\it intrinsic}) colour, specific star formation rates (SSFR) and stellar mass history through redshift. This is achieved using the information stored in the publicly available \eagle\ database \citep{McAlpine2016}. We first identify all galaxies that by $z=0$ are quenched, based on their position in either the $u^{*}-r^{*}$- or SSFR-stellar mass planes. We then compute the difference between the lookback time at which each passive galaxy was last part of the star-forming population, and when they subsequently appeared in the quenched population (based on these two different methodologies, which we describe in more detail below). Note that in either of our definitions we take into account that galaxies in the corresponding planes evolve and thus, the exact definition of star-forming and passive are redshift dependent. 

\begin{figure}
\centering
\includegraphics[width=01\columnwidth]{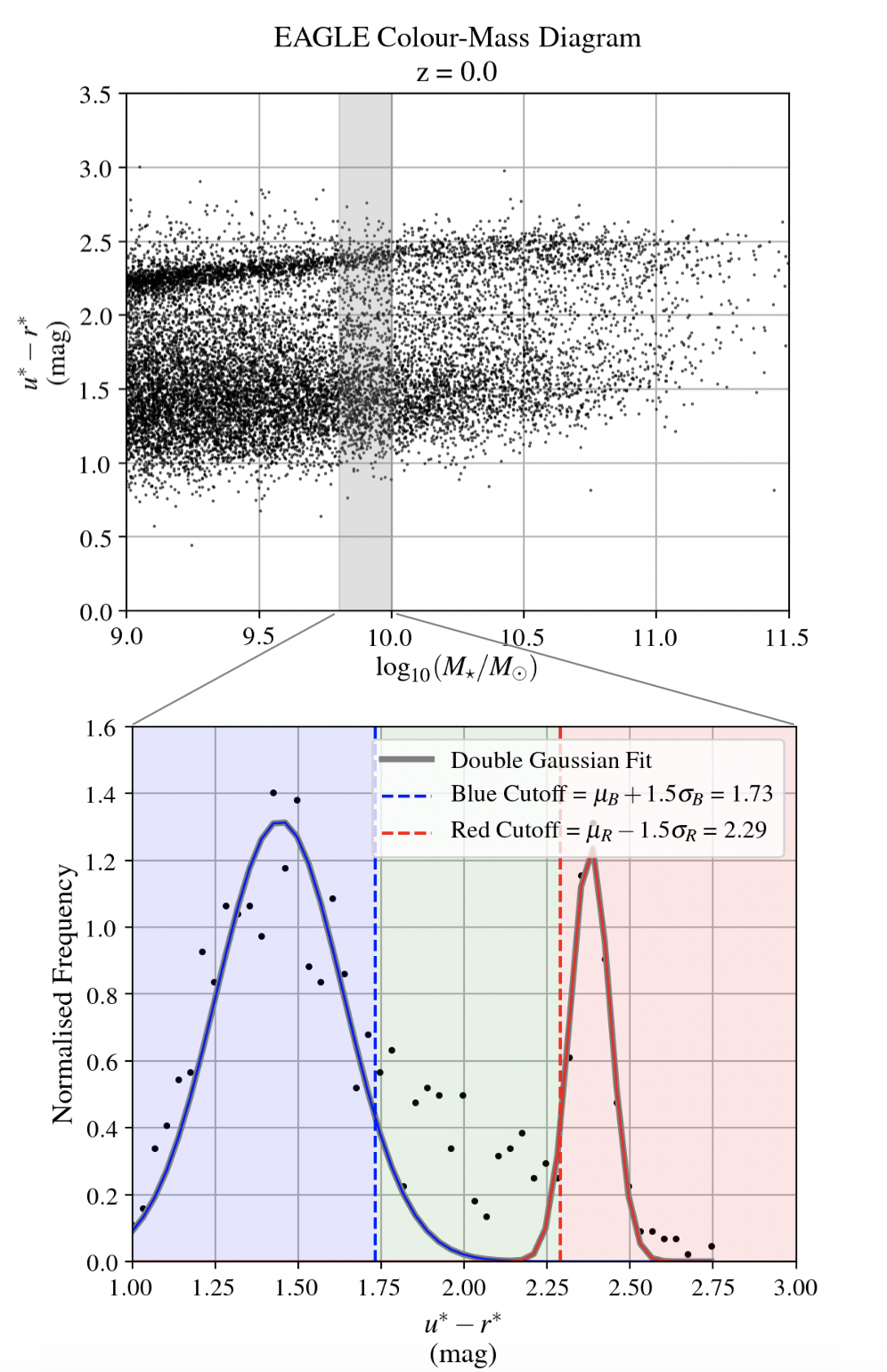}
\caption{An illustration of the Gaussian fitting criteria to identify the red sequence and blue cloud in binned stellar mass space that we apply to \eagle. Each population was defined by its mean ${\pm} 1.5{\sigma}$. The region between these two populations defines the green valley (see $\S$~\ref{colorselec} for details).}
\label{fig:Gaussian}
\end{figure}
\begin{figure}
\centering
\includegraphics[width=01\columnwidth]{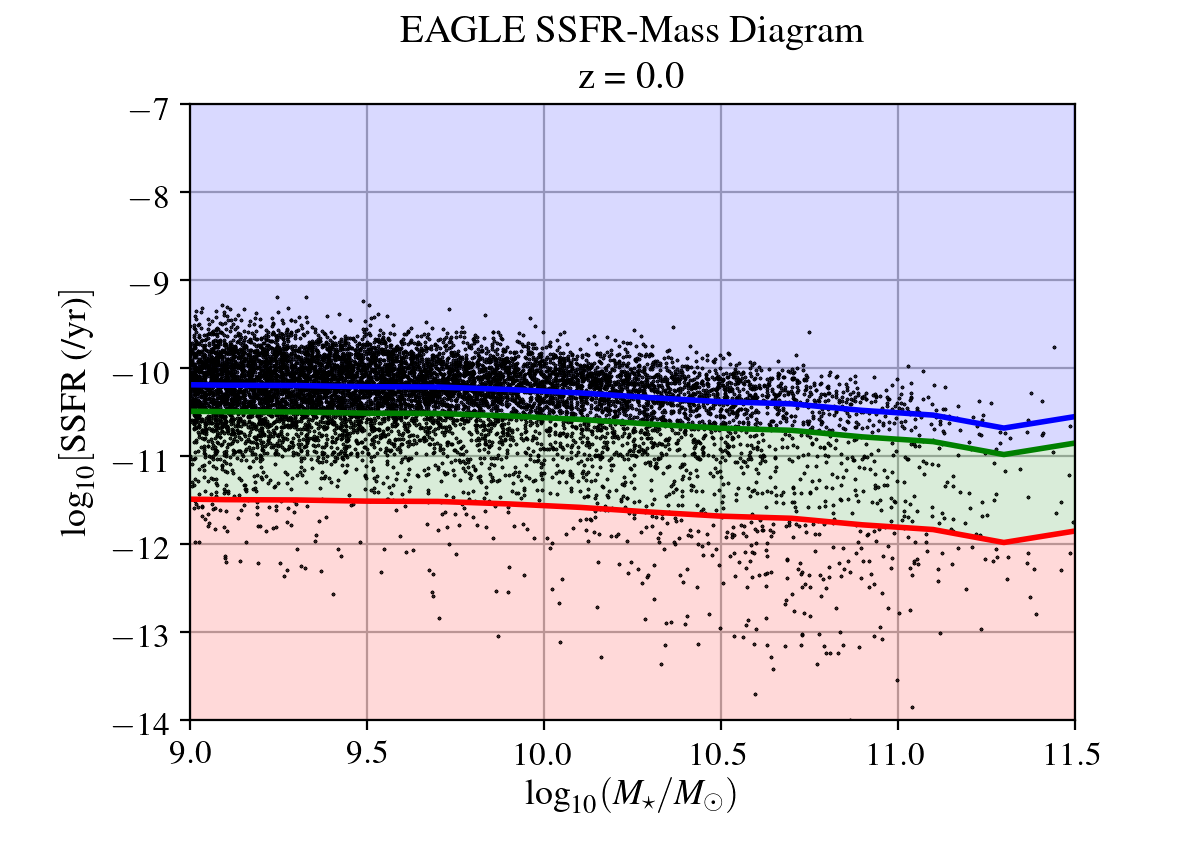}
\caption{An illustration of the SSFR-$M_{\star}$ criteria for the red sequence and blue cloud in binned stellar mass space. Each population was defined by cutoffs in terms of the position of the main sequence at each stellar mass bin, with values $c_{\rm low} = 0.05$ (red line) and $c_{\rm high}=0.5$ (green line), below, in between and above these values, galaxies are classified as passive, transitioning and star-forming, respectively (see $\S$~\ref{sfrselec} for details).}
\label{fig:SSFR_Def}
\end{figure}

\subsubsection{Color definition of quenching timescale}\label{colorselec}

We use the $(u^{*} - r^{*})-M_{\star}$ plane to create our primary definition (definition 1, referred to as `D1' hereafter) of quenching timescales, similar to the work of \citet{TRAYFORD2016} and \citet{Nelson2018}.  To  categorise  whether  a  galaxy  was part of the blue cloud, red sequence, or green valley, a double-Gaussian distribution was fit to the colour histogram of \eagle\ galaxies in $15$ bins of stellar mass from $10^9 M_{\odot} - 10^{12} M_{\odot}$, at each snapshot. The means and standard deviations in $(u^{*} - r^{*})$ colour of the blue cloud and red sequence were estimated and accordingly fitted  with  a  least  squares  curve  fitting  function  in  SciPy. We set limits on the allowable mean of the red sequence and blue cloud, the bounds being $\pm 30\%$ of the expected mean colour of each population. We adopt values for the expected mean of the red sequence and blue cloud which we define in a similar fashion to \citealt{TRAYFORD2016}, namely: $(u^*-r^*)_{\rm \ red,\ exp}=0.2{\log}_{10}(M_{\star}/M_{\odot})-0.25z^{0.6}+0.4$ and $(u^*-r^*)_{\rm \ blue,\ exp}=0.2{\log}_{10}(M_{\star}/M_{\odot})-0.25z^{0.6}-0.5$. We place no strict priors on the {$\sigma$} of each population, other than ensuring that its value in one stellar mass bin does not vary by more than 50\% from its adjacent stellar mass bins. Using  these  values,  we  define  a  galaxy  to  be  quenched  if $1.5\sigma$ above the red-Gaussian mean in a given $M_{\star}$ bin (and snapshot), and star-forming if $1.5\sigma$ below the blue-Gaussian mean colour in a given $M_{\star}$ bin (and snapshot), according to Eqs.~\ref{passdef1}~and~\ref{passdef2},  respectively.  Using  this  definition  at $z=0$, we  identify  $3,054$  quenched  galaxies,  $2,871$  galaxies  in  the process of quenching, and $7,273$ star-forming galaxies. This selection process is illustrated in Fig.~\ref{fig:Gaussian}.

\begin{align}
(u^{*} - r^{*})_{\textrm{\ blue, cutoff}} &= {\mu}_{{\rm u^{*} - r^{*}}\textrm{, blue}}+1.5{\sigma}_{{\rm u^{*} - r^{*}}\textrm{, blue}}
\label{passdef1}\\ (u^{*} - r^{*})_{\textrm{\ red, cutoff}} &= {\mu}_{\rm {u^{*} -r^{*}}\textrm{, red}}-1.5{\sigma}_{{\rm u^{*} - r^{*}}\textrm{, red}}\label{passdef2}
\end{align}

\subsubsection{SSFR-stellar mass definition of quenching timescale}\label{sfrselec}

The secondary definition (definition 2, referred to as `D2' hereafter) for classifying galaxies as star forming or quenched used their SSFRs directly, instead of using galaxy colour as a proxy of star formation activity. The SSFR-$M_{\star}$ plane produces a less well defined population bimodality, which made producing the cutoff definitions somewhat subjective \citep{WYDER07,Ciambur2013,Davies2016GAMAENVIRONMENT,Davies18b_sub}. We include both D1 and D2 for completeness, and remarkably find our conclusions to be largely consistent between the two definitions. 

The method we adopt for D2 is based on the position of the main sequence star formation, which we quantify with the mean SSFR of galaxies with a star formation cutoff defined in \citet{Furlong2015}: ${\rm log_{10}}({\rm SSFR}_{\rm MS}/{\rm yr^{-1}}) >-11+0.5z$.  We took the mean SSFR of the star forming main sequence (MS), $\langle \rm SSFR_{\rm MS}(M_{\star})\rangle$, using 15 bins in stellar mass, spanning from $M_{\star}=10^9M_{\odot}$ to $M_{\star}=10^{12}M_{\odot}$. We then consider `transitioning' galaxies as those which have SSFRs between $\rm c_{\rm low}\,\rm \langle SSFR(M_{\star})\rangle$ and $\rm c_{\rm high}\,\rm \langle SSFR(M_{\star})\rangle$. Star-forming and passive galaxies are consequently defined as those with SSFRs $>\rm c_{\rm high}\,\rm \langle SSFR(M_{\star})\rangle$ and $< \rm c_{\rm low}\,\rm \langle SSFR(M_{\star})\rangle$, respectively. We adopt values for $\rm c_{\rm low}$ and $\rm c_{\rm high}$ such that the number of passive, transitioning and star-forming galaxies is similar to those obtained with the color-stellar mass selection of $\S$~\ref{colorselec}, in an attempt to match the subjective thresholds and populations over different definitions. As such, we find the most appropriate multiplicative values to be 
$\rm c_{\rm low} = 0.05$ and $c_{\rm high}=0.5$. Using this definition at $z=0$, we obtain $3,368$ quenched galaxies, $2,226$ quenching galaxies, and $7,613$ star-forming galaxies. This selection process is illustrated in Fig. \ref{fig:SSFR_Def}.  Comparing these two definitions, we find that $2,990$ galaxies are identified as quenched at $z=0$ for both D1 and D2 (97.9\% of D1 galaxies appear in D2, while 88.8\% of D2 galaxies appear in D1). Observational papers have adopted values such as $c_{\rm low}=0.025-0.5$ \citep{Bethermin15,Wang18,Davies18c_sub}, and thus, we consider the values adopted here reasonable compared to those in observations.

\subsection{Reconstructing galaxy formation histories}\label{sec:histories}

To reconstruct the formation histories of galaxies, we use the public database of the \eagle\ simulations, described in \citet{McAlpine2016}. Below, we describe the procedure we follow. We limit our study to galaxies with stellar masses  $M_{\star} >10^9\,\rm M_{\odot}$, that are considered as converged \citep{SCHAYE2015}. 

\begin{figure*}
\includegraphics[width=0.93\textwidth]{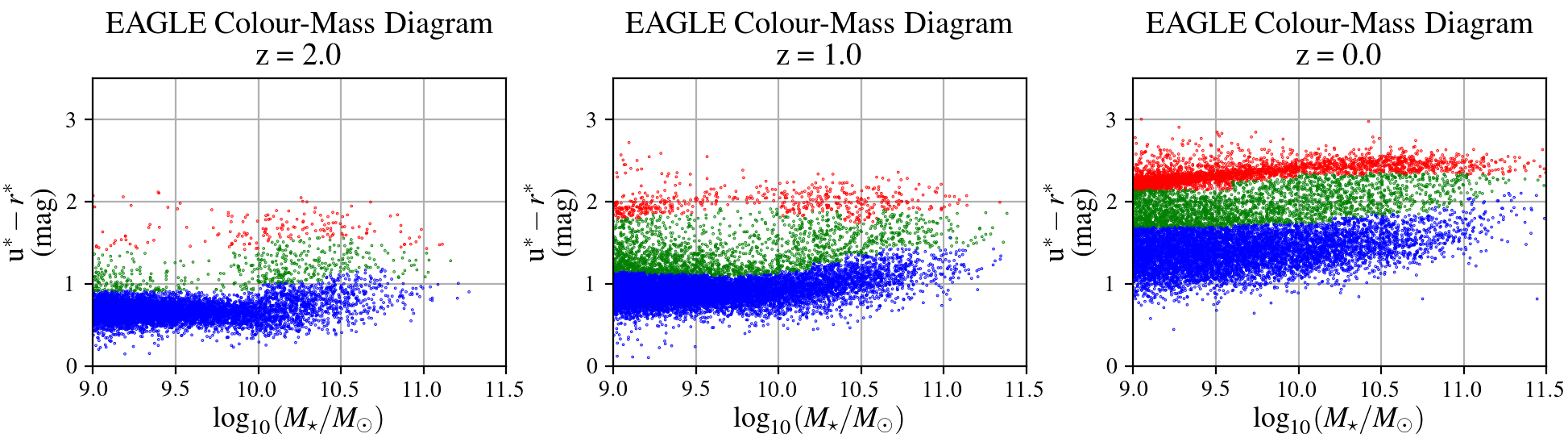}
\caption{The red sequence developing in \eagle\ from $z=2$ to $z=0$, as labelled. We see a scarcity of quenched galaxies at intermediate stellar mass, where galaxies are not massive enough to host a large central SMBH, but are also too massive to be greatly influenced by environmental factors, in concurrence with the work of \citet{Davies18c_sub} from GAMA observations, and others. Colours indicate star formation status as per $\S$~\ref{colorselec}: blue symbols show star-forming galaxies, red symbols show quenched galaxies (part of the red sequence), and green symbols show galaxies in the green valley.}
\label{fig:RSEvolution}
\end{figure*}

\begin{figure*}
\includegraphics[width=0.93\textwidth]{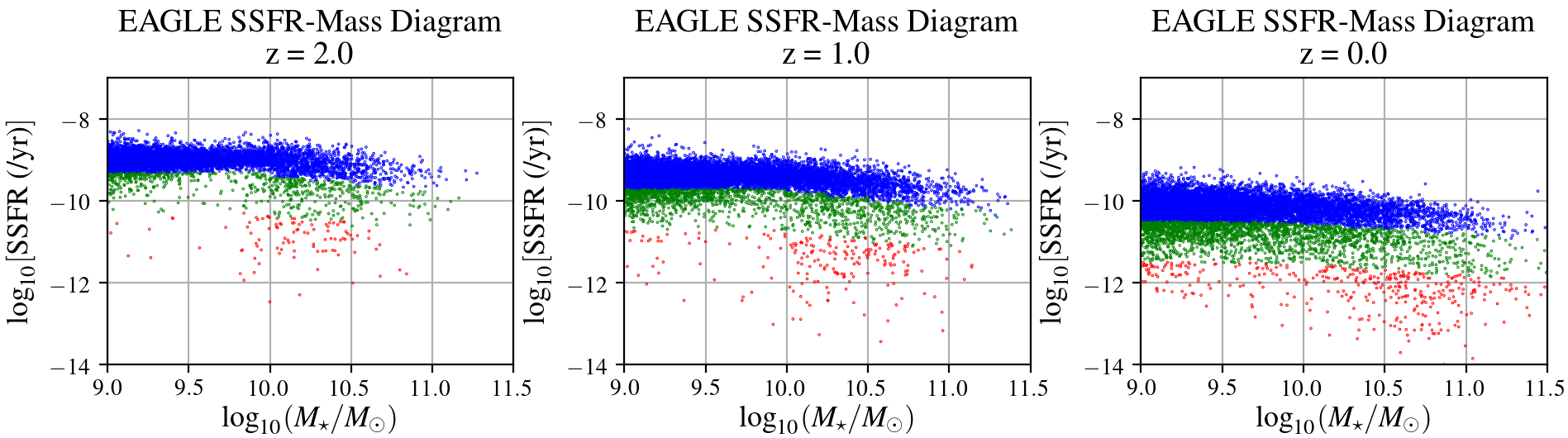}
\caption{Distribution of galaxies in the SSFR-$M_{\star}$ plane, evolving from $z=2$ to $z=0$, as labelled, in \eagle. The main sequence of star formation decreases in SSFR over time while more galaxies enter the green transition population moving towards the passive population. The passive population develops from both stellar mass ends with a slight under-density at intermediate stellar mass, similar to Fig.~\ref{fig:RSEvolution}, in concurrence with the work of \citep{Davies18c_sub} from GAMA observations, and others. Colours indicate star formation status as per $\S$~\ref{sfrselec}: blue symbols show star-forming galaxy (member of main sequence of star formation), red symbols show quenched galaxies (member of the passive cloud), and green symbols show transitioning galaxies.}
\label{fig:MSEvolution}
\end{figure*}

For all galaxies at $z = 0$, we define their status as a star-forming, quenching, or passive galaxy based on D1 and D2. Quenched galaxies at $z=0$ are identified and tracked back through redshift. This yields a sample size of $n = 3,054$ galaxies for D1 and $n = 3,368$ galaxies for D2.  The $3,054$ and $3,368$ galaxies, respectively, are followed using the merger tree information stored in the database, described in \citet{Qu17}. If multiple progenitor galaxies exist, the most massive (in stellar mass) progenitor in the previous snapshot is selected as the main progenitor, and we continue tracking this branch. Only redshifts below $z = 2$ are considered, since at higher redshifts, the passive population is seldom developed, as demonstrated in Fig.~\ref{fig:RSEvolution} for D1 and Fig. \ref{fig:MSEvolution} for D2. At each snapshot, we record the $30$~pkpc stellar mass, $30$~pkpc gas mass, $30$~pkpc colour, $30$~pkpc SFR, $30$~pkpc SMBH mass, full subhalo gas mass, FOF group mass and status of central or satellite of each of the $z = 0$ quenched galaxies which are being tracked back.

To determine whether a galaxy was central or satellite at the time it left the blue cloud, we took the SubGroupNumber of the galaxy at the snapshot it left the blue cloud, and for each snapshot before and after. To be considered a central galaxy, a galaxy was required have ${\rm SubGroupNumber}=0$ at the snapshot before, during {\it and} after it left the star-forming population. Those galaxies not classed as central according to this definition were classified as satellites. Using 3 snapshots as opposed to 1 to consider for classification avoids some identification issues with the FOF algorithm in EAGLE - those galaxies falling into a larger halo may not yet be considered a satellite of this system, while they are actually being physically affected by the larger group \citep{TRAYFORD2016}. Using 3 snapshots instead of 1 for the satellite definition yields $2,419$ satellites as opposed to $2,263$ for D1 - meaning $156$ galaxies or about $6$\% of satellites could be potential ``splashback'' galaxies. Since these galaxies have been processed in a group environment, we believe this classification is physically accurate.

In addition to the quantities above, we also calculate the cumulative number of galaxy mergers, and net SMBH accretion rate.  The cumulative number of mergers is calculated by recording the number of times a galaxy had multiple progenitors in the range $z=2$ to $z=0$, requiring that the secondary-to-primary galaxy stellar mass ratio be no larger than 10. We remind the reader that each progenitor galaxy was required to be above the stellar mass threshold of $M_{\star}=10^9M_{\odot}$. The net specific SMBH accretion rate was calculated based on the change in SMBH mass within over 4 snapshots about the time the galaxy left the blue cloud, divided by relevant lookback time difference between the snapshots concerned and the galaxy's stellar mass. We included multiple snapshots to calculate the SMBH accretion rate to avoid stochasticity significantly affecting the derived values. We also tested other definitions of SMBH accretion rates, such as the amount of mass the SMBH grew while the galaxy was transitioning the green valley, divided by that timescale, or the dynamical timescale of the central $3$~kpc stellar overdensity. We found our conclusions to be robust under these different definitions (see $\S$~\ref{agneffectsec}). 

Once each of the mentioned quantities was calculated, we processed the history of each galaxy to determine which snapshot the galaxy was:

\begin{enumerate}
\item{First considered quenched for D1 \& D2, and:}
\item{Last considered star-forming for D1 \& D2.}
\end{enumerate}

The difference between the lookback times of these snapshots was taken as the quenching timescale of the given galaxy: {\it the time it took for the galaxy to traverse the transition region}, for D1 and D2, respectively. Note that because we are using snapshots to define quenching timescales, our measurements are limited to the cadence of the snapshots in \eagle, which prevents us from measuring quenching timescales much smaller than $1$~Gyr (unless quenching occurred at higher redshift). In principle, the quenching could have occurred at any point in the snapshot adjacent (in the appropriate temporal direction) to the one recorded defining the lookback times of when the galaxy joined the red sequence or left the blue cloud. To account for this potential discrepancy, we calculated the quenching timescale of each galaxy 100 times with red sequence and blue cloud lookback times calculated with added/subtracted random proportions of the appropriate adjacent snapshot cadence. We used all 100 samples of quenching timescales for each definition to generate error margins for our quenching timescale distributions.

\begin{figure}
\centering
\includegraphics[width=01.05\columnwidth]{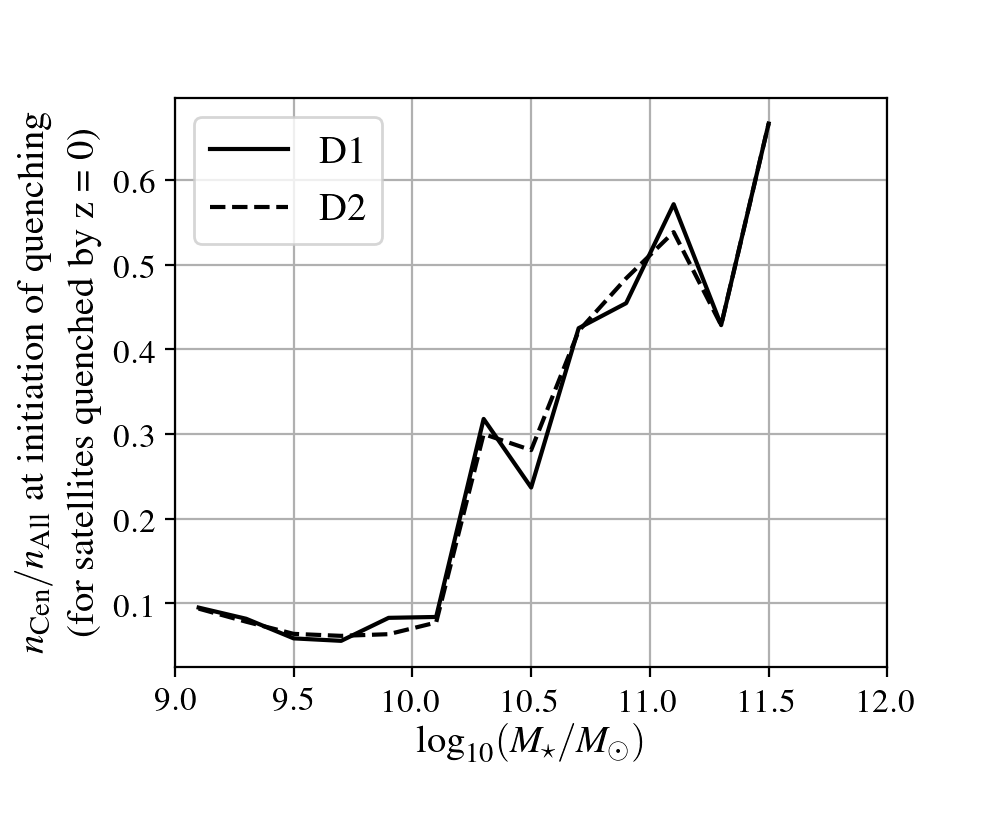}
\caption{The proportion of quenched satellite galaxies at $z=0$, which were classed as central around the time they left the blue cloud, as a function of stellar mass. The proportion increases significantly with stellar mass for both D1 (solid line) and D2 (dashed line) definitions, indicating that the physics that quenched the high-mass satellite galaxies is likely the same as that of massive central galaxies.}
\label{fig:SatCenFractions}
\end{figure}

\begin{figure*}
\centering
\includegraphics[width=1\textwidth]{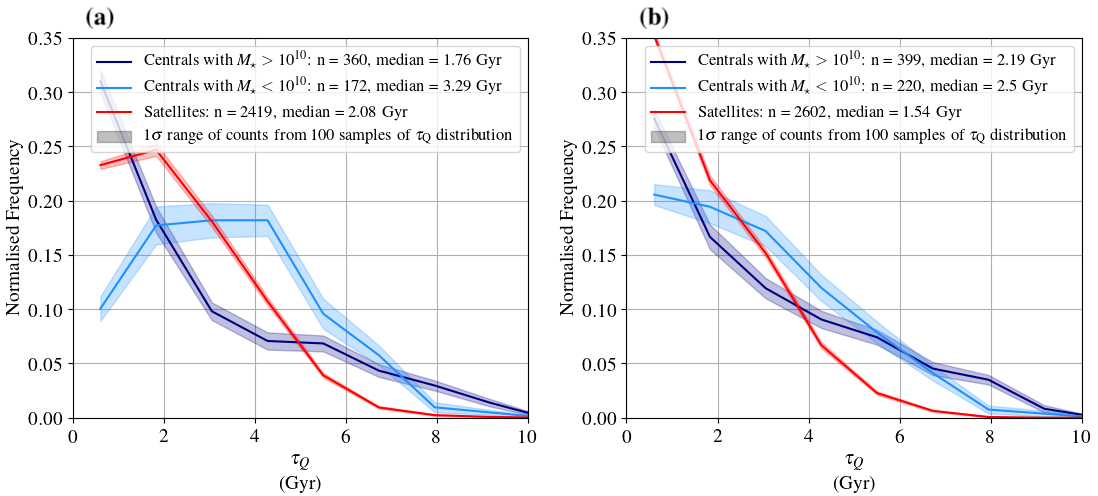}
\caption{ The quenching timescale (${\tau}_{\rm Q}$) distribution of $z=0$ passive galaxies, separated into whether they were massive centrals (blue, $M_{\star}>10^{10}M_{\odot}$), smaller centrals (cyan, $M_{\star}<10^{10}M_{\odot}$) or satellite galaxies (red) around the time they left the star-forming population, based on D1 (panel (a)) and D2 (panel (b)) definitions of galaxy state. Shaded regions represent $1\sigma$ error margins on the frequency in each bin from the 100 $\tau_{\rm Q}$ samples, each of which have also been jackknifed. In general, satellite galaxies have a higher frequency of $\tau_{\rm Q}\approx 2-3$~Gyr compared to massive centrals for both D1 and D2. The less massive central group exhibits a convincing peak in quenching timescales at ${\tau}_{\rm Q} = 3-4$~Gyr for D1, and a less prominent but noticeable increase in frequency at ${\tau}_{\rm Q} = 3-4$~Gyr for D2.}
\label{fig:1_SM_SSFR_Comp}
\end{figure*}

\section{Quenching timescales of galaxies and their physical origin}\label{quenchingsec}

\begin{figure*}
\centering
\includegraphics[width=01.0\textwidth]{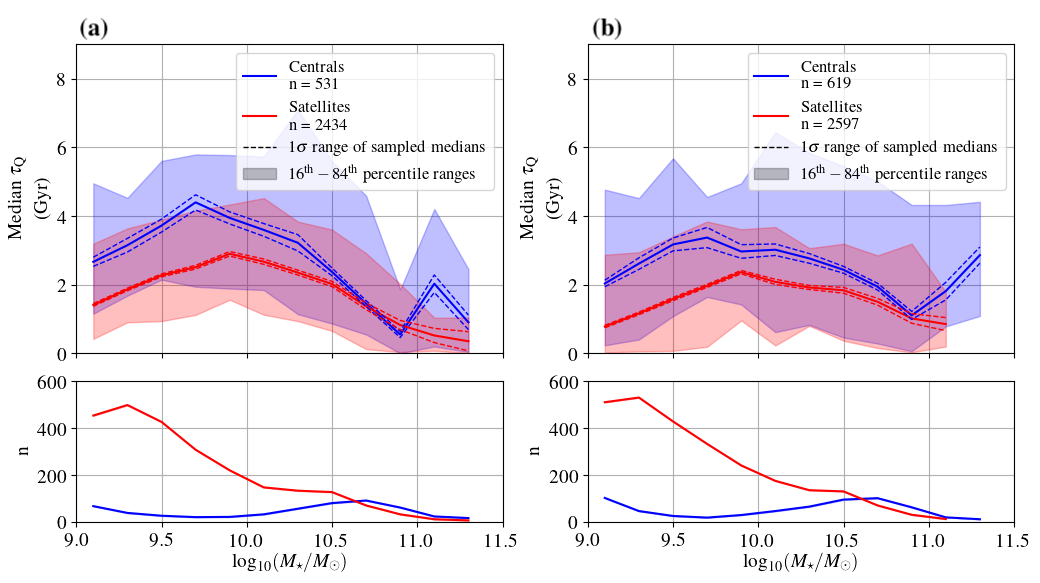}
\caption{Median quenching timescales of passive galaxies at $z=0$, separating those that were central (blue) and satellite (red) galaxies around the time they left the star-forming population, as a function of stellar mass, for the D1 (panel (a)) and D2 (panel (b)) definitions of green valley. The shaded regions corresponds to the $16^{\rm th}-84^{\rm th}$ percentile range for ${\tau}_{\rm Q}$ in each stellar mass bin for satellites and centrals, while the dotted margins represent the $1\sigma$ error margin on the median in each bin after considering 100 ${\tau}_{\rm Q}$ samples and jackknifing each. A peak in quenching timescale is observed at $M_{\star} \approx 10^{9.7}\,\rm M_{\odot}$ for all galaxies. A positive correlation between stellar mass and quenching timescale is observed at lower stellar masses, while a negative correlation is observed at higher stellar masses. This behavior is seen in both centrals and satellites.}
\label{fig:2ab_SM_SSFR}
\end{figure*}
\begin{figure}
\centering
\includegraphics[width=1\columnwidth]{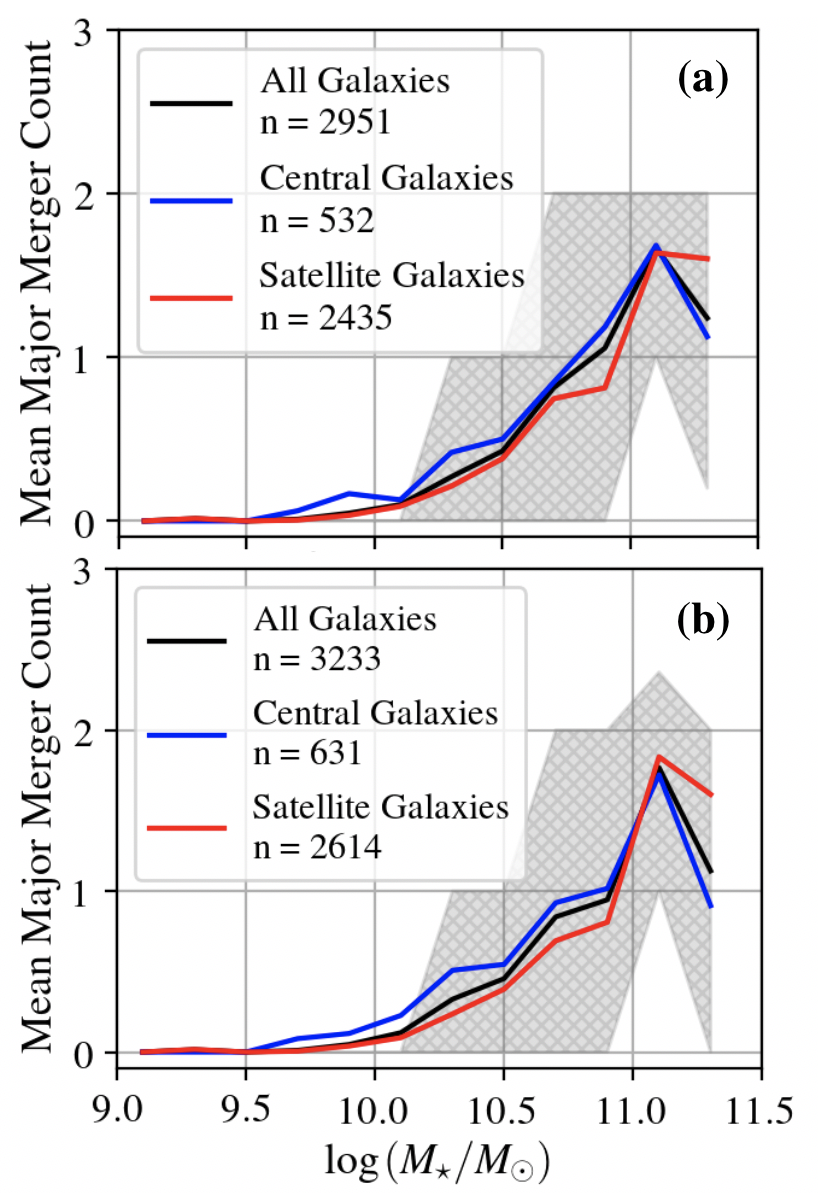}
\caption{The mean number of galaxy mergers for passive galaxies at $z=0$ by stellar mass bin, separated into whether they were central or satellite around the time they left the star-forming population. We show this for definitions D1 (panel (a)) and D2 (panel (b)) as a function of stellar mass, only including mergers with mass ratios below 10. The shaded regions correspond to the $16^{\rm th}$-$84^{\rm th}$ percentile of total merger count in each stellar mass bin (not distinguished between satellite and central galaxies). We seldom observe mergers for low stellar mass galaxies, however at $M_{\star}=10^{10.5}\,\rm M_{\odot}$ the mean merger count in each bin picks up and increases rapidly with stellar mass.}
\label{fig:2c_SM_SSFR}
\end{figure}

Here we present our analysis of galaxy quenching timescales in probing the physics behind star formation quenching, and ultimately galaxy bimodality. The results to follow include quenching timescale distributions, with populations separated based on their condition of central or satellite galaxy, as well as other properties, which we identify with certain mechanisms of quenching. One should note that the exact values of the quenching timescales are sensitive to the exact definition of green valley/transitioning region. However, we generally find that the trends presented here are robust to these different definitions.  All quantities (unless otherwise stated) are taken at the snapshot the given galaxy left the star-forming population, in order to connect to the physical origin of the quenching. In addition, the classification of central or satellite galaxy is taken at the time the galaxy left the star-forming population. This distinction is especially important for passive satellite galaxies at $z=0$, as a significant fraction of them were centrals at the time they left the star-forming sequence (see Fig.~\ref{fig:SatCenFractions}).

\subsection{Quenching timescales of central and satellite galaxies}
Fig.~\ref{fig:1_SM_SSFR_Comp} shows the quenching timescale distribution produced by \eagle\ galaxies based on the two definitions D1 and D2 outlined in $\S$~\ref{colorselec} and $\S$~\ref{sfrselec}, respectively, split further into low-mass centrals, high-mass centrals, and satellite galaxies. The low stellar mass to high stellar mass threshold for central galaxies was set at $M_{\star}=10^{9.75}M_{\odot}$, so as to split the stellar mass bimodality for centrals which is illustrated in Fig.~\ref{fig:2ab_SM_SSFR}. We find that the D2 definition gives slightly lower quenching timescales compared to the results obtained by the D1 definition, which we expect since even in the case of a sudden cessation of star formation, colour transformation would require some non-zero timescale to complete. Nonetheles, the differences in ${\tau}_{\rm Q}$ distributions found between samples are present for both D1 and D2. The main difference is seen for low mass centrals, where for D1 the ${\tau}_{\rm Q}$ peak is observed at $3-4$~Gyr, while for D2 there is a distinct knee (as opposed to a peak) in the distribution, a symptom of the systematically reduced quenching timescales we see for D2. Overall, the median quenching timescales for all selections shown fall in the $1.4$~Gyr - $3.5$~Gyr range, which agrees well with the measurements reported by \citet{Nelson2018} using the Illustris-TNG simulations. 

We interpret the low-mass central population to represent central galaxies existing in lower mass halos. For these galaxies, much longer $3$-$4$~Gyr timescales are commonplace. Since we do not expect environmental effects or AGN activity to be significant for these galaxies (see $\S$~\ref{secenvirons} and $\S$~\ref{agneffectsec}), it is clear that the main mode of star formation quenching in these galaxies is stellar feedback, which appears to act over significantly longer timescales compared to the typical massive central or satellite galaxy. For higher-mass central and satellite galaxies (existing in higher mass halos), we note that multiple factors are expected to be affecting the quenching of these galaxies, and it is therefore more difficult to interpret these histograms without further decoupling. This said, the reader should note that satellites do exhibit a knee in their distribution at the $3$~Gyr mark when compared to high-mass central galaxies, for both D1 and D2 definitions of quenching timescale.

\citet{TRAYFORD2016} studied the pathways in the colour-magnitude diagram of galaxies, and found that low-mass satellites and massive centrals quench first, followed by intermediate mass centrals and satellites. Here, we see that the mechanisms behind those different pathways for centrals and satellites leave a clear imprint in the quenching timescales of these two populations. Hence, in the following subsections, we explore in more detail what drives the distribution of quenching timescales, and in which circumstances the quenching timescales of galaxies can elongate or shorten. 

\subsection{The dependence of the quenching timescale on stellar mass}\label{sec:tq-stellarmass}

Fig.~\ref{fig:2ab_SM_SSFR} shows how quenching timescales depend on stellar mass in \eagle, with these timescales calculated using D1 (panel (a)) and D2 (panel (b)) quenching definitions. D1 and D2 give mostly consistent results in terms of distribution shape. We also split galaxies into satellites and centrals, as per $\S$~\ref{sec:histories}. For both centrals and satellites, we see that there is a peak in quenching timescales at $M_{\star}=10^{9.7}\,\rm M_{\odot}$, which corresponds well with the under-density of quenched galaxies at intermediate stellar mass documented in \citet{TRAYFORD2016} and Fig.~\ref{fig:RSEvolution}, as well as the observational findings of \citet{Davies18c_sub}. As we will discuss in detail, we find that the dearth of galaxies at that mass scale is directly a consequence of their quenching timescales being on average longer than galaxies at other mass scales, regardless of whether a galaxy is classified central or satellite. 

For central galaxies, we see their quenching timescales to be systematically longer than satellites below $10^{10}\,\rm M_{\odot}$, meaning the peak at intermediate stellar mass is less defined for centrals. This corresponds to the lower mass central population in Fig. \ref{fig:1_SM_SSFR_Comp}. The quenching timescales of satellites and centrals appear to converge above this stellar mass threshold, indicating that the physics occurring to quench high mass galaxies is not a function of central/satellite classification (and thus less likely to be linked to galaxy environment). The turning point at intermediate stellar mass, however, suggests that there are two possibly mechanisms acting for satellites and centrals at the lower stellar mass end, while there appears to be a different, but unique mechanism responsible for quenching high-mass galaxies. The turning point represents the stellar mass at which gas can accrete and stay in a galaxy most efficiently, where gas content in the host halo is most rich. This is further investigated in $\S$~\ref{secenvirons}, in our discussion of gas fractions. 

The bottom panels of Fig.~\ref{fig:2ab_SM_SSFR} show the histogram of stellar masses of centrals and satellites galaxies. The frequency of satellite galaxies decreases steadily with increasing stellar mass, while centrals display a double peaked distribution. This double-peaked distribution is connected to the efficiency of baryon collapse and star formation being maximal at stellar masses $\approx 10^{10}\,\rm M_{\odot}$ in both simulations and observations \citep{Behroozi13,Eckert17}.

To explain the double-peaked distribution of the central galaxy stellar mass, and also to probe the mechanisms behind quenching timescales, we explore the connection between the cumulative number of major galaxy mergers in galaxies and the previously mentioned mass transition scale of $\approx 10^{10}\,\rm M_{\odot}$ in Fig.~\ref{fig:2c_SM_SSFR}. We remind the reader that the mass ratio of each merger was required to be less than 10 for consideration. \eagle\ shows a dramatic increase in the number of mergers experienced by galaxies at $M_{\star}\gtrsim 10^{10}\,\rm M_{\odot}$, which closely corresponds to the turning point in quenching timescales. This also agrees with observation-based findings, where an increase in the incidence of mergers has been found as stellar mass increases \citep{Robotham14}. Above this transition mass, the galaxy population in each stellar mass bin moves from satellite- to central-dominated, and we expect from disk- to bulge- dominated: high stellar mass galaxies which have been quenched by $z=0$ will typically have undergone multiple merging events. We did not conduct an in-depth investigation on morphology induced changes in quenching timescales. \citet{Correa18} presents a detailed analysis of the relationship between morphological and color transformation of galaxies in \eagle, and found that the timescale for galaxies to change their color (similar to our $\tau_{\rm Q}$ in the case of definition D1) were very weakly correlated. They do, however, find a strong correlation for the time at which galaxies join the red sequence and the morphology of galaxies, with early-types typically having joined much earlier than disk galaxies. Furthermore, \citet{Tacchella19} used the IllustrisTNG simulations to demonstrate a similar result: that galaxies do not necessarily change their morphology as they transition through the green valley. Instead, a galaxy's morphology is largely set while the galaxy is still star-forming. Both of these studies align well with the observational work of \citet{Cortese19}, who find using the SAMI Galaxy Survey that the quenching of star formation in satellite galaxies does not necessarily imply that  kinematic transformation has occurred. They attribute the discrepancy in spin observed between local satellite and central galaxies to the fact that satellites develop their angular momentum at a slower rate compared to centrals after entering more massive halos.

For the rest of this paper, we focus on investigating in depth the physical drivers behind the complex $\tau_{\rm Q}-M_{\star}$ relationship we find for central and satellite galaxies.

\subsection{External quenching mechanisms and the connection with quenching timescales}\label{secenvirons}
\begin{figure*}
\centering
\includegraphics[width=0.75\textwidth]{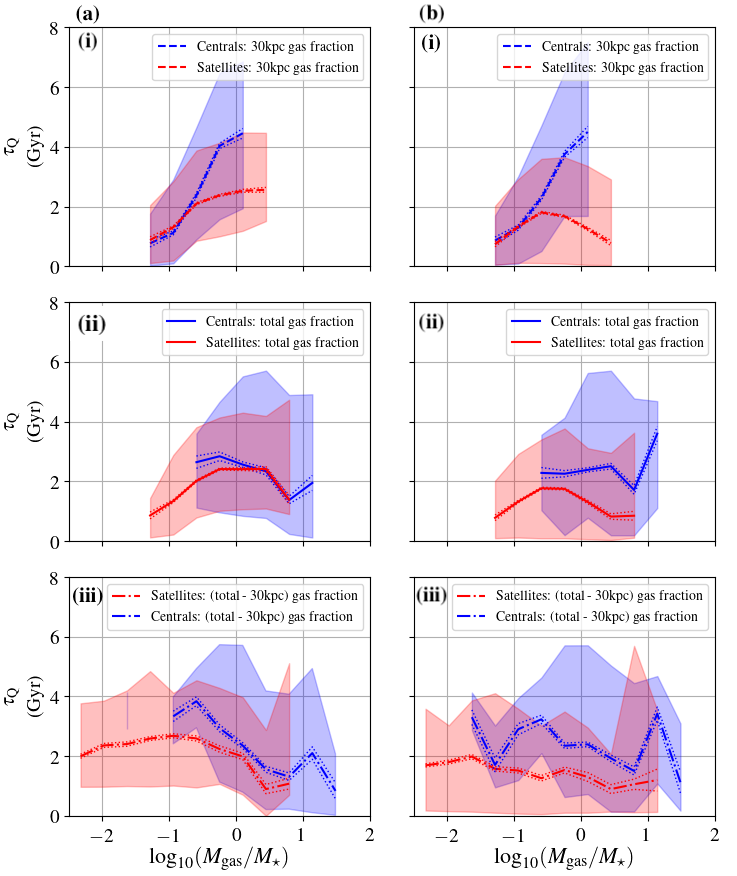}
\caption{The median quenching timescale of $z=0$ passive galaxies, separated into whether they were centrals (blue) or satellites (red) around the time they left the star-forming population, for the D1 (panel (a)) and D2 (panel (b)) definitions of green valley, as a function of various individual galaxy gas fractions. The gas fractions are normalised by the stellar mass of the galaxy. We define ``gas mass'' using three different criteria: total gas mass in the subhalo (solid lines), gas within the central $30$~kpc (dashed lines), and the outer gas mass (total gas mass subtracting the gas mass within $30$~kpc; dash-dot lines). Shaded regions correspond to the $16^{\rm th}-84^{\rm th}$ percentile range of quenching timescales fraction in each gas fraction bin, while dotted lines about the medians correspond to the $1\sigma$ range on the median quenching timescale in each bin calculated from the 100 ${\tau}_{\rm Q}$ samples, each of which was also jackknifed.}
\label{fig:GGF_TQ}
\end{figure*}
\begin{figure*}
\centering
\includegraphics[width=1.0\textwidth]{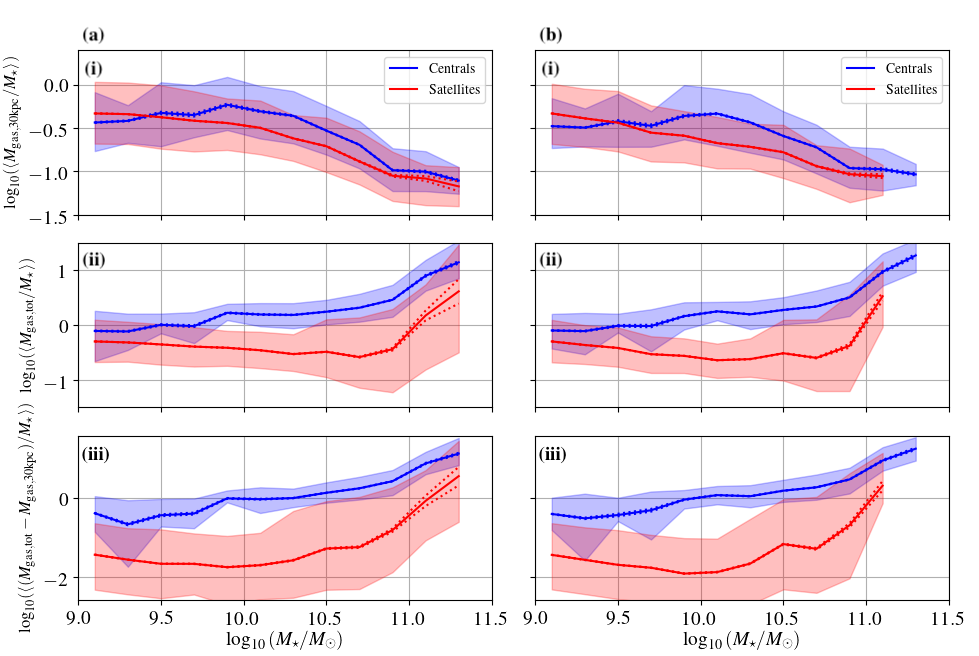}
\caption{The median gas fraction (defined as gas mass normalised by stellar mass) of $z=0$ passive galaxies as a function of stellar mass, separating them by whether they were centrals (blue) or satellites (red) around the time they left the star-forming population, for the D1 (panel (a)) and D2 (panel (b)) definitions of green valley. We show the same gas fractions as in Fig.~\ref{fig:GGF_TQ} as a function of stellar mass. Shaded regions correspond to the $16^{\rm th}-84^{\rm th}$ percentile range of gas fractions in each stellar mass bin, while dotted lines correspond to the $1\sigma$ error on the median gas fractions in each bin from the jackknifed samples.}
\label{fig:SM_GGF}
\end{figure*}

In this section, we explore the effect of environment on the quenching timescales of both central and satellite passive galaxies. First, we consider the role of the galaxy gas content in star formation quenching, both strongly linked in the ``bathtub model'' \citep{Dave12,Lilly13,Schawinski2014} of galaxy gas inflow and outflow. In this model, any quenching mechanism acts by depleting a galaxy's ISM gas reservoir in some way. Before considering which specific gas-removal mechanisms may be at work, we first look at how the gas content of galaxies and their host halos relates to star formation quenching timescales in \eagle. \citet{Lagos2015Gas} and \citet{Crain2017} investigated the numerical convergence of the the \eagle\ simulations and corresponding ${\rm H}_{2}$ and HI gas results respectively. \citet{Lagos2015Gas} find that the ${\rm H}_{2}$ content of \eagle\ galaxies is unaffected by resolution for galaxies with ${\rm M}_{{\rm H}_{2}}>10^{7.5}\,\rm M_{\odot}$, while \citet{Crain2017} find that \eagle\ systematically underestimates the HI column density of galaxies with stellar masses $<10^{10}\,\rm M_{\odot}$. The results we present only depend on the relative gas masses of galaxies meaning this systematic effect is less important for our conclusions.

Fig.~\ref{fig:GGF_TQ} shows the quenching timescale as a function of the gas fraction of galaxies using 3 different definitions: taking the gas mass within a spherical aperture of $30$~kpc (row (i); gas ``inside'' galaxies); taking the total gas mass of the subhalo ($M_{\rm Gas, Total}$; row (ii)); and taking the gas mass in the subhalo {\it outside} the $30$~kpc spherical aperture (i.e. subtracting the gas ``inside'' galaxies from the total, $M_{\rm Gas, Total}-(M_{\rm Gas,30kpc})$;  row(iii)). These are normalised by the stellar mass of the galaxy. One could argue that the demarcation of gas ``inside'' and ``outside'' galaxies at $30$~kpc is arbitrary; however when using $50$~kpc, $70$~kpc and $100$~kpc apertures as the threshold radii, we find the results to be consistent to those presented below. We show our results for both centrals and satellites that by $z=0$ are classified as passive according to $\S$~\ref{sec:definitions} - D1 in column (a), D2 in column (b). We remind the reader that these gas fractions and stellar masses are calculated at the time these galaxies left the star-forming population, so as to connect our results to the physical mechanisms which induced star-formation quenching.

Exclusively considering the gas within a $30$~kpc radius of a galaxy (row (i) of Fig. \ref{fig:GGF_TQ}) we find that there is a strong positive correlation between quenching timescales and the $30$~kpc gas fraction of central galaxies over the full span of $M_{\rm Gas, 30kpc}/M_{\star}$ values, from $10^{-1.5}$ to $10^{0.5}$ using both D1 and D2 ${\tau}_{\rm Q}$ definitions. Central galaxies with $M_{\rm Gas, 30kpc}/M_{\star}\lesssim 0.1$ show quenching timescales of $\approx 1$~Gyr on average, while the highest $30$~kpc gas fractions have quenching timescales of $>4$~Gyr on average: very much in the tail of the overall histogram in Fig.~\ref{fig:1_SM_SSFR_Comp}. This is an indication that the gas ``inside'' galaxies is more important for up-regulating star formation (or more specifically, elongating the time that it takes to cease) than the gross total gas mass within a SubHalo, which is likely dominated by hot, diffuse halo gas. \citet{Lagos2015Gas} showed that in \eagle\ most of the dense, cool gas is within a $30$~kpc aperture, while \citet{Mitchell2018} showed that most of that gas is rotationally supported. Hence, it is safe to assume that the gas within $30$~kpc we are analysing here is ISM gas. Comparitively, satellite galaxies show a different trend between $\tau_{\rm Q}$ and the $30$~kpc gas fractions. For D1, $\tau_{\rm Q}$ increases with gas-fraction similarly to centrals up to $(M_{\rm Gas, 30kpc})/M_{\star}\approx10^{-0.5}$; however, above this scale $\tau_{\rm Q}$ flattens at around $2.5$~Gyr. For D2, $\tau_{\rm Q}$ of satellites increases up to $(M_{\rm Gas, 30kpc})/M_{\star}\approx10^{-0.5}$, above which the dependence reverses and quenching timescales drop. It is possible that the $30$~kpc radius encompasses more virialised gas as opposed to ISM gas for satellites when compared to centrals, especially at high $M_{\rm Gas, 30kpc}$ values. If this hot halo gas has the effect of stripping galaxies rather than feeding star formation, it would offer an explanation to this drop/plateau in quenching timescales. 

For both central and satellite galaxies (and for both definitions of green valley, D1 and D2) we see that the total gas fraction (solid lines) and outer gas fraction (dot-dash lines) are weakly correlated with $\tau_{\rm Q}$. This is, again, due to the dual effect of hot halo gas being capable of stripping galaxies of their gas through ram pressure stripping, while cooled ISM gas can concurrently offer a gas reservoir for accretion to galaxies and eventual star formation. 

We continue our investigation by probing how the aforementioned gas fractions vary as a function of stellar mass for galaxies in Fig.~\ref{fig:SM_GGF}, with the aim of interpreting the peak of quenching timescales at intermediate masses discussed in $\S$~\ref{sec:tq-stellarmass}. Presented is the gas fraction for those galaxies which were quenched at $z=0$ (according to definitions D1 and D2). We note, however, that including the full \eagle\ galaxy population at $z=0$ leads to similar results. We find that the total gas fraction and outer gas fraction for centrals to be systematically higher than satellites, indicating that centrals on average have more total and virialised gas assigned to their SubHalo; while for the $30$~kpc halo gas fraction, we find that centrals and satellites have fairly similar gas content, with centrals being slightly more gas rich within $30$~kpc at higher stellar masses. This is consistent with the observations of \citet{Brown2017}, using atomic hydrogen spectral stacking of SDSS galaxies. If we consider the total gas mass fractions and ``outer'' gas fraction, each stays steady with increasing stellar mass in both D1 and D2 up to $M_{\star}\approx 10^{10}M_{\odot}$, where each increase with stellar mass (most steeply for satellites). In the case of the gas ``inside'' galaxies, we instead find for D1 and D2 that centrals exhibit a noticeable peak in gas content ($M_{\rm Gas,30kpc}/M_{\star}\approx0.5$) at intermediate stellar masses, $\approx 10^{9.7}-10^{10.3}\,\rm M_{\odot}$, in a stellar mass regime incrementally larger, by up to $\approx 0.5$~dex, than that of the peak of the quenching timescales at $M_{\star}=10^{9.7}M_{\odot}$. Above $M_{\star}\approx 10^{10.3}\,\rm M_{\odot}$, inner gas fractions of centrals start to decrease with increasing stellar mass. For satellites, $30$~kpc gas fractions appear to decrease monotonically with stellar mass, starting at approximately the same value as centrals at $M_{\star}=10^{9}M_{\odot}$, indicating that their increased stellar mass content does not necessarily bring with it more gas. As we will show in $\S$~\ref{agneffectsec}, AGN activity at stellar masses above the group inner gas fraction peak starts to be efficient enough to keep the gas halos hot and also eject gas from galaxies.

\begin{figure*}
\centering
\includegraphics[width=1\textwidth]{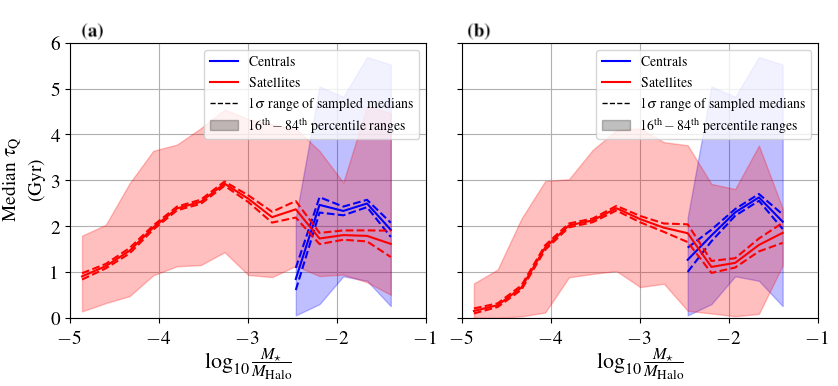}
\caption{The median quenching timescale of the $z=0$ passive galaxies, separating them into whether they were centrals (blue) or satellites (red) around the time they left the star-forming population, for the D1 (panel (a)) and D2 (panel (b)) definitions, as a function of the galaxy-halo mass ratio: $M_{{\star}}/M_{\textrm{Halo}}$. We define this quantity by using the host halo mass and stellar mass of the galaxies at the redshift when the galaxy left the star-forming population. Shaded regions correspond to the $16^{\rm th}-84^{\rm th}$ percentile of ${\tau}_{\rm Q}$ in each stellar-halo mass ratio bin, while dotted lines about the medians correspond to  $1\sigma$ range of median mass fractions calculated from the jackknifed samples. Galaxies more massive relative to their halo typically exhibit longer quenching timescales, this trend being especially obvious for satellite galaxies.}

\label{fig:4_SM_SSFR}
\end{figure*}

\begin{figure*}
\centering
\includegraphics[width=1\textwidth]{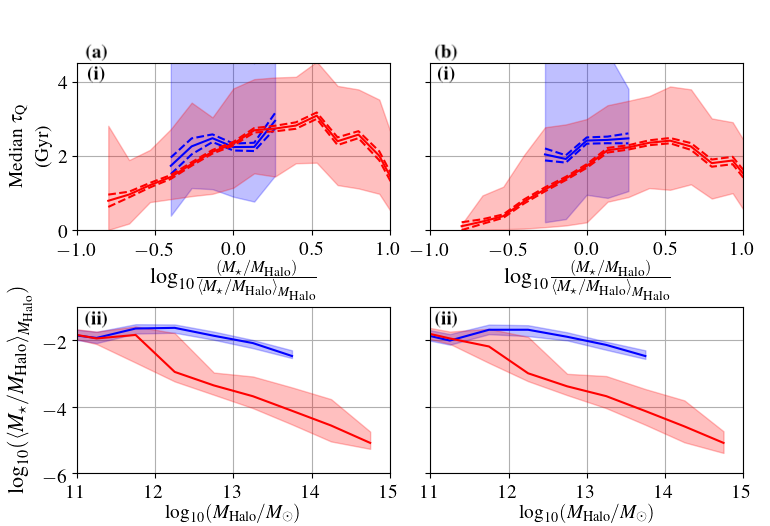}
\caption{The median quenching timescale of the $z=0$ passive galaxies, separating them into whether they were centrals (blue) or satellites (red) around the time they left the star-forming population, for the D1 (column a) and D2 (column b) definitions, as a function of the galaxy-halo mass ratio {\it excess}: $\frac{M_{{\star}}/M_{\textrm{Halo}}}{\langle M_{{\star}}/M_{\textrm{Halo}}\rangle_{M_{\rm Halo}}}$. We show the halo mass - stellar-halo mass ratio relation in row (ii). We define this quantity by using the host halo mass and stellar mass of the galaxies at the redshift when the galaxy left the star-forming population. Shaded regions correspond to the $16^{\rm th}-84^{\rm th}$ percentile of ${\tau}_{\rm Q}$ in each stellar-halo mass ratio bin, while dotted lines about the medians correspond to  $1\sigma$ range of median mass fractions calculated from the jackknifed samples. Galaxies more massive relative to their halo average stellar-halo mass ratio, for a given halo mass, typically exhibit longer quenching timescales, this trend being especially obvious for satellite galaxies.}
\label{fig:4b_SM_SSFR}
\end{figure*}

Now, we investigate how the mass of a galaxy in comparison to its host halo can influence star formation quenching. On average, we would expect relatively small galaxies to be more prone to environmental effects from neighbouring large galaxies in the same host halo \citep{Peng2012,Davies18c_sub}. This is because low mass galaxies lack the self-gravity needed to compensate for the pressure of the intra-halo gas medium, and are therefore likely to be stripped of their star-forming gas. Estimates of the quenching timescales of dwarf satellite galaxies with stellar masses below $10^9M_{\star}$ indicate longer timescales as the mass of the satellite increases \citep{Fillingham15}. {\citet{Wetzel2013} find that the quenching timescales of more massive satellites galaxies ($M_{\star}>10^{9.5}\,\rm M_{\odot}$) tend to decrease with stellar mass due to group pre-processing}. To explore the effect of relative galaxy mass on quenching timescales in \eagle, we elect to use the fraction $M_{\star}/M_{\textrm{Halo}}$, with $M_{\star}$ and $M_{\rm Halo}$ being the stellar mass and group mass of the host halo in which the galaxy resided before leaving the star-forming population, to probe whether satellite galaxies small relative to their halo are quenched over different timescales compared to more massive satellites. In addition, we extend the investigation to include galaxies that were centrals the time they left the star-forming population. 

Fig.~\ref{fig:4_SM_SSFR} clearly shows, for both D1 and D2, that the quenching timescales of satellite galaxies are positively correlated with the $M_{{\star}}/M_{\textrm{Halo}}$ ratio for lower stellar-halo mass ratios below $M_{\star}/M_{\rm Halo}=10^{-3}$. Above this point, quenching timescales decrease. One should note that there are very few satellite galaxies above this mass ratio, and those that do exist are more likely to behave like centrals (as shown in $\S$~\ref{sec:tq-stellarmass}). For central galaxies, the $M_{{\star}}/M_{\textrm{halo}}$ ratio spans a much narrower range, as expected from the tight stellar-halo mass relation of central galaxies in \eagle\ \citep{SCHAYE2015,Guo2016}. It appears for both D1 and D2 that quenching timescales increase for central galaxies which are more massive relative to their halo, before decreasing at the highest mass ratio bin for D1, at $M_{\star}/M_{\rm Halo}\approx10^{-2}$. Short quenching timescales, $\lesssim 1$~Gyr dominate the satellite galaxies that are less massive compared to their host halo, while stellar mass-to-halo mass ratios in excess of $10^{-3.5}$ tend to be associated to quenching timescales $\gtrsim 2$~Gyr. We remind the reader that for this investigation we imposed a stellar mass floor of $10^9\,\rm M_{\odot}$, and as such the range of $M_{{\star}}/M_{\textrm{Halo}}$ is likely limited on the lower end.

To demonstrate that our conclusions are not influenced by our restricted halo mass sample, we also investigate how the quenching timescales of central and satellite galaxies vary as a function of stellar-halo mass {\it excess} - i.e. the stellar-halo mass ratio relative to the median value of the sample at fixed halo mass. We calculate this excess by computing the median stellar-halo mass ratio for centrals and satellites in bins of halo mass (row (ii) in Fig. \ref{fig:4b_SM_SSFR}) and then calculate, for each galaxy and its given host halo, whether it was more massive than average (a separate calculation is performed for centrals and satellites). Our results in row (i) of Fig.~\ref{fig:4b_SM_SSFR} show for satellites (which span a large range in stellar-halo mass ratio excess) that their quenching timescales increase with stellar-halo mass ratio excess to a peak of $3$~Gyr up to a ratio excess value of $\approx 10^{0.5}$ for both D1 (panel (a)) and D2 (panel (b)), and then decrease with increasing mass ratio excess. In other words, despite a restricted halo mass sample, if a satellite galaxy is less massive relative to its halo than we would expect {\it given its host halo mass}, then it is more likely to quench over shorter timescales. Satellites with a higher stellar-halo mass excess above this turning point correspond to the more massive galaxies which have enough AGN activity as to quench star formation on shorter timescales, see $\S$~\ref{agneffectsec}. In this regime, it appears that they behave similar to central galaxies. In comparison, the quenching timescales of central galaxies increase with stellar-halo mass ratio excess monotonically over a smaller dynamic range. 

To further ensure our conclusions are not are not driven by our resolution-induced stellar mass cut, we have also checked that the results for satellite galaxies displayed in Figs. \ref{fig:4_SM_SSFR} and \ref{fig:4b_SM_SSFR} are robust when examining only a sub-sample of halos with a mass $\ge 10^{13}\,\rm M_{\odot}$. In this regime, stellar-to-halo mass ratios $\ge 10^{-4}$ are complete, and we find that the trends discussed for satellite galaxies remain consistent. Based on this evidence, we can conclude that in \eagle, smaller satellite galaxies in larger halos tend to quench more quickly on average than larger galaxies which are more dominant in their halo. Physically, one could argue that this effect is likely due to the stripping of gas in these smaller galaxies from intra-halo gas pressure, acting to quickly reduce star formation rates and trigger the colour transformation. These conclusions are consistent against D1 and D2 definitions (panel (a) and panel (b) respectively). Note that both the increasing gas fraction with stellar mass and the implied larger stellar-to-halo mass ratio with increasing stellar mass for satellites work towards increasing the quenching timescale with increasing stellar mass in this population. Thus, the trend seen in Fig.~\ref{fig:2ab_SM_SSFR} at $M_{\star}\lesssim 10^{10}\,\rm M_{\odot}$ is well explained by the environmental trends reported here. The stellar mass turning point in satellites occurs directly before the ``inner'' galaxy gas fraction starts to decrease (Fig.~\ref{fig:SM_GGF}). 

Despite the small dynamic range in $M_{\star}/M_{\rm Halo}$ obtained for central galaxies in \eagle, we see that there is a tendency for the quenching timescale to increase with the stellar-to-halo mass ratio, which is more evident for the D2 definition of green valley. This suggests that the scatter of the stellar-halo mass relation is related to how quickly central galaxies quench at fixed halo mass, which in turn correlates with how gas-rich a halo is, and how much SMBH accretion a galaxy has experienced (see discussion in $\S$~\ref{agneffectsec}). Future surveys probing the HI content of galaxies and halos in the local Universe will be able to unveil any systematic effect of the gas fraction of halos on the halo-stellar mass relation and the frequency of green valley galaxies. Those surveys include those to be carried out by the Australian and South African Square Kilometre Array \citep{Blyth2015}.

Recently, \citet{Davies18c_sub} showed, using the GAMA survey \citep{Driver09}, that the quenched fraction of satellite galaxies at fixed halo mass increases steeply with the ratio of central-to-satellite stellar mass. They defined quenched galaxies with a method analogous to D2 (using SSFR as the indicator of galaxy transformation). Because the stellar mass of central galaxies is well correlated with the host halo mass, one would expect a similar correlation to exist between the passive fraction and the host halo-to-satellite mass ratio. That result naturally fits with our findings with \eagle\ and our interpretations, where quenching timescales significantly lengthen for satellites that are massive with respect to their host halo. The latter is expected to have an impact on the overall passive fraction in a way that there are less passive satellites of high stellar-to-host halo mass ratios. The results of \citet{Davies18c_sub} therefore provide observational evidence for our findings here.

\begin{figure*}
\centering
\includegraphics[width=01.0\textwidth]{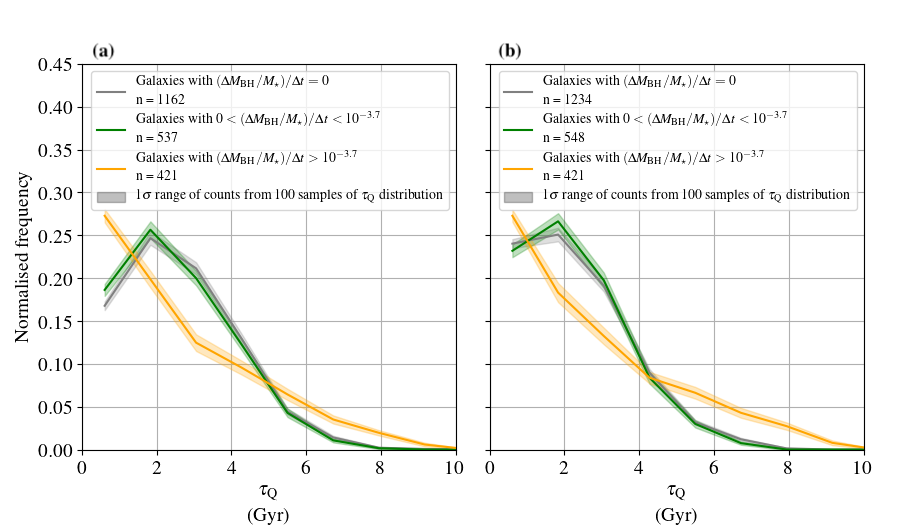}
\caption{The quenching timescale distribution for all galaxies in bins of zero SMBH accretion rate (grey) vs low- and high-SMBH accretion rate galaxies (yellow and green respectively) for both D1 (panel (a)) and D2 (panel (b)). Shaded regions represent $1\sigma$ error margins of frequency in each bin from the 100 $\tau_{\rm Q}$ samples, each of which have also been jackknifed. We elect to not split the sample for satellites and centrals since we found their behaviour with black hole activity to be identical. For low and high-SMBH accretion rate galaxies (non-zero), the result appear similar. The zero-accretion rate sample and low-accretion rate sample in both D1 and D2 exhibit a more spread ${\tau}_{\rm Q}$ distribution, with D1 low accretion rate galaxies actually exhibiting a peak in quenching timescale frequency at $\approx 2$~Gyr instead of the lowest bin. The high-accretion rate sample, in contrast, shows a clear peak at the lowest quenching timescale bin and decreases rapidly at longer quenching timescales for both D1 and D2. For both definitions, a galaxy with a zero or low SMBH accretion rate is much more likely to have a quenching timescale of $3-6$~Gyr than a galaxy which does exhibit black hole accretion.}
\label{fig:9_SM_SSFR}
\end{figure*}

\subsection{AGN activity and quenching timescales}\label{agneffectsec}

\begin{figure*}
\centering
\includegraphics[width=01.0\textwidth]{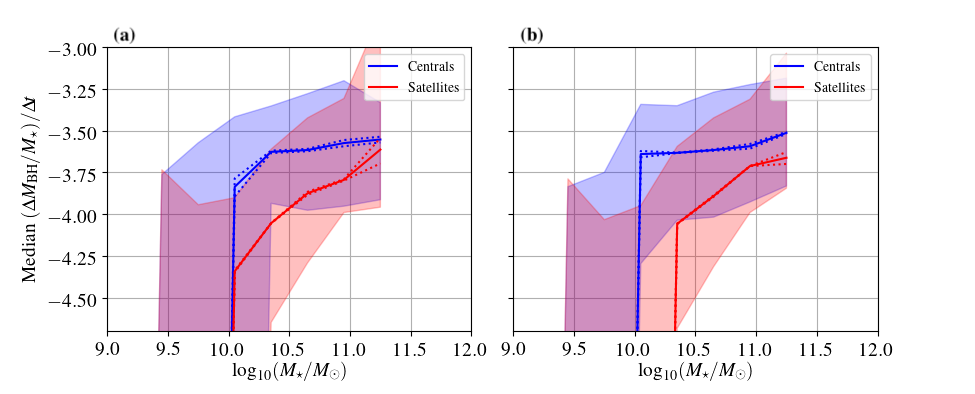}
\caption{The median SMBH accretion rate in bins of stellar mass for passive galaxies at $z=0$, separated into whether they were satellites (red) or centrals (blue) around the time they left the star-forming population. Panel (a) uses D1, while panel (b) uses D2. The shaded regions correspond to the $16^{\rm th}-84^{\rm th}$ percentile range of SMBH accretion rate in each stellar mass bin, while dotted lines correspond to the $1\sigma$ error on calculating the median from a jackknifed sample of the galaxies in each bin. SMBH activity noticeably picks up for both centrals and satellites above $10^{10} M_{\odot}$, the stellar mass regime where we expect satellites and centrals to be quenched by a similar mechanism based on Fig. \ref{fig:2ab_SM_SSFR}.}
\label{fig:13a_SM_SSFR}
\end{figure*}

AGN are known to be capable of quenching star formation due to both heating and ejection of star-forming gas material (see \citealt{Fabian12} for a review on observational evidence of AGN feedback and \citealt{Bower17} for a detailed analysis of AGN feedback in \eagle). The AGN heating effect in \eagle\ depends on how quickly a SMBH can accumulate the necessary energy to heat gas particles to the required temperature (see $\S$~\ref{subgridphysics}). Thus, AGN feedback in \eagle\ critically depends on the SMBH accretion rate. 

Fig.~\ref{fig:9_SM_SSFR} shows how the net SMBH accretion rate (normalised by galaxy stellar mass) influences the quenching timescales of central galaxies that at $z=0$ are passive, for both definitions D1 and D2. We split the galaxy population into $3$ bins of net accretion rate: one for galaxies showing zero-accretion rate, one for those of low-accretion rate, and one for those of high-accretion rate. We distinguish between the latter two samples with a SMBH accretion rate cutoff at the median of non-zero accretion rate galaxies, $\frac{{\Delta}(M_{\rm BH}/M_{\star}}{\Delta t})=10^{-3.7}$. Roughly half of all galaxies would exhibit a non-zero accretion rate, the rest split roughly evenly between the two non-zero accretion rate samples. {Zero accretion rates are likely a consequence of the mass resolution of {\sc eagle}. Physically, we simply consider these galaxies in this category of very low accretion rates}. We remind the reader that the net SMBH accretion rate is calculated as the change in SMBH mass the galaxy experienced over 4 snapshots, around the time the galaxy departed from the star forming population to the passive population. This change in mass was then divided by its stellar mass and by the lookback time difference between the two relevant snapshots. 

For both D1 \& D2, we find that the zero- and low- SMBH accretion rate galaxies (grey and green lines respectively) exhibit a peak in quenching timescales at ${\tau}_{\rm Q}\approx 2$~Gyr; while the higher accretion rate galaxies (orange lines) show a much more defined peak at quenching timescales $\lesssim 1$~Gyr. This indicates that a higher SMBH accretion rate (and corresponding luminosity) could be acting to shorten quenching timescales through the efficient heating and ejection of inter-stellar gas. This is in line with the findings of \citet{TRAYFORD2016} in \eagle, which show that central galaxies with overly massive SMBH are more likely to be passive compared to those with less massive SMBHs. {This also aligns well with the observational results of \citet{HerreraCamus2019}, who show that AGN-driven outflows are capable of expelling central molecular gas on characteristic timescales of $\approx 0.2$~Gyr}. According to our results, the galaxies with massive SMBHs would be more likely to quench faster, and consequently become part of the red sequence. { We do see, however, a small amount of galaxies with high SMBH accretion rates which exhibit longer quenching timescales, up to $\approx 8$~Gyr, indicating that not all galaxies hosting bright AGN will quench rapidly. Rather, it appears that these galaxies with active AGN are less likely than others to quench on timescales of $\gtrsim 2-3$~Gyr, the timescale which characterises the rest of the population. }

Recall Fig.~\ref{fig:2ab_SM_SSFR}, where central and satellite galaxies display a peak in quenching timescales at around $10^{9.7}\rm \, M_{\odot}$ in \eagle. To relate back, we now show how the net SMBH accretion rate depends on stellar mass for central and satellite galaxies in Fig.~\ref{fig:13a_SM_SSFR}. Zero SMBH accretion rate galaxies were included in this context, and accordingly, below $M_{\star}=10^{10}\,\rm M_{\odot}$, \eagle\ galaxies display a steep drop-off in AGN activity. Our findings agree with the work of \citet{Bower17}, who found that the SMBH accretion rates of galaxies in \eagle\ increase rapidly in halos of masses above $10^{12}\,\rm M_{\odot}$. {In the L100 {\sc eagle} run on  average, satellites in a $10^{12}M_{\odot}$ halo have a stellar mass of $\approx 10^{9.7}M_{\odot}$, while centrals in a $10^{12}M_{\odot}$ halo have stellar mass ~$\approx 10^{10.3}M_{\odot}$}. This corresponds to the intermediate stellar mass range where we find the rapid increase in AGN activity. This tells us that AGN activity is only likely to affect quenching timescales in galaxies which have higher stellar mass, $M_{\star}\gtrsim 10^{10}\,\rm M_{\odot}$. AGN activity picks up rapidly at this stellar mass value, and provides further evidence to the conclusions we make in $\S$~\ref{secenvirons}. Thus, we conclude that the turning point in quenching timescale dependence with stellar mass is, at least in part, due to AGN activity - which heats the interstellar gas and removes it from the galaxy while keeping the halo gas hot. {Some galaxies with significant AGN activity could still be capable of holding onto their cool interstellar gas reservoir (depending on AGN type and accretion rates), illustrated by the tail in the high-accretion rate galaxy sample in Fig.~\ref{fig:9_SM_SSFR}, while some show a definite shortening of their quenching timescale to $<1$~Gyr.} 

Interestingly, massive satellite galaxies in \eagle\ that are passive by $z=0$ also exhibit an increase in their net SMBH accretion rate at high stellar masses, which indicates that AGN activity is aiding the quenching of massive satellite galaxies. The latter is necessary to make these galaxies passive, since their environment alone is not enough to drive the cessation of star formation in these galaxies. Also note that many central galaxies that quenched due to the effect of AGN feedback end up being satellites by $z=0$. This fact was illustrated for D1 and D2 in Fig~\ref{fig:SatCenFractions}.

\begin{figure*}
\centering
\includegraphics[width=01.0\textwidth]{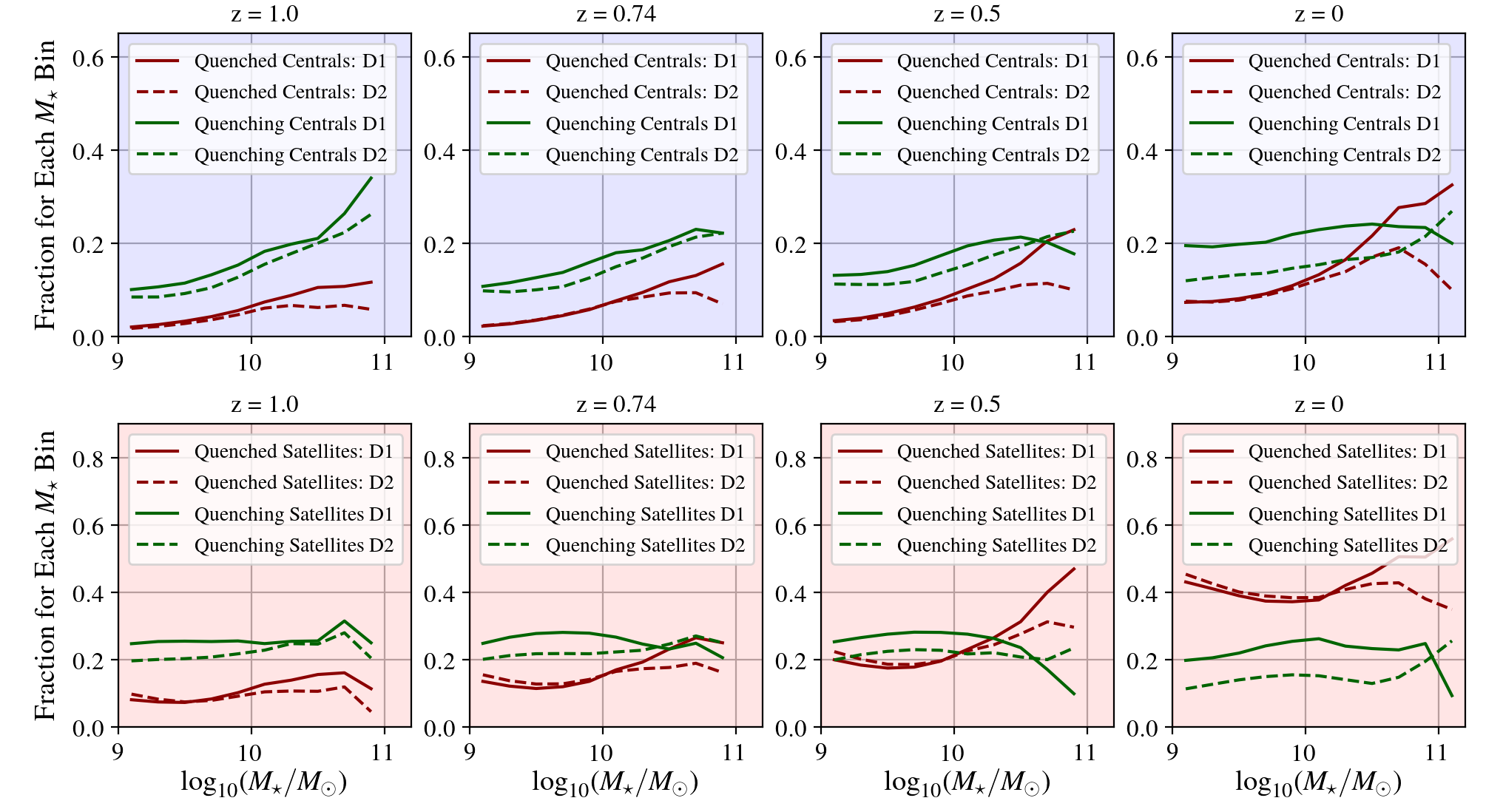}
\caption{The fraction of galaxies (as a function of stellar mass) that are part of the quenching population or passive population in \eagle\ based on D1 and D2, for redshifts between $z=1$ and $z=0$. Over time, for all stellar masses, the quenched fraction of the population increases (corresponding to where galaxies leave the quenching population). This first occurs most notably in the higher stellar mass extremity. At $z=1$ we see a transition population fraction be highest in the higher stellar mass bins, and over redshift, these galaxies transverse to the quenched population, leaving a lower fraction of transition galaxies at higher stellar mass at $z=0$.}
\label{fig:Fraction}
\end{figure*}

\subsection{Predictions for the green valley across time}\label{sec:predictions}

In coming years, highly complete surveys focused on intermediate-redshift galaxies such as DEVILS \citep{Davies2018DEVILS}, could offer insight into the properties of transitioning green-valley galaxies as a function of group properties and redshift. We present here in Fig.~\ref{fig:Fraction} the proportion of galaxies in \eagle\ that are part of the green valley/ transition region or red sequence/ passive population, based on D1 and D2. 

As expected, the overall fraction of passive galaxies increases with time (red lines in Fig.~\ref{fig:Fraction}), but the fraction of green valley galaxies evolves very weakly with time. This is related to the average timescale for this transitions being rather short. However, some interesting differential behaviour is seen. One of them is that the dependence of the fraction of green valley galaxies with stellar mass for centrals and for definition D1, which at $z=1$ is positively correlated, systematically flattens towards $z=0$, even displaying an inversion of the relation at stellar masses $\gtrsim 10^{10.5}\,\rm M_{\odot}$ and at $z\lesssim 0.5$. Interestingly, the fraction of central green valley galaxies in the colour-magnitude diagram at intermediate redshifts ($z\approx 0.5$) peaks at stellar masses of $\approx 10^{10.2}\,\rm M_{\odot}$, due to the quenching timescales being longer at this stellar mass. The flattening is also seen for centrals and the D2 definition, but is less significant than that seen for D1. 

For satellites, we also see a rapid increase of the fraction of passive galaxies, which is more striking than that seen for centrals. Interestingly, the green valley fraction of satellites in definition D1 is close to constant at $25$\% at $M_{\star}\lesssim 10^{10.2}\,\rm M_{\odot}$, with the fraction at higher stellar masses decreasing with time. The inflection stellar mass is due to the mass scale above which we see the quenching timescale quickly decreasing with increasing stellar mass in satellites (see. Fig.~\ref{fig:2ab_SM_SSFR}). Note that this inflection stellar mass also corresponds to when we see a minimum in the fraction of passive satellites, which is clear in both definitions, particularly at intermediate redshifts, $z\approx 0.5-0.75$. For definition D2, we see again a rather flat fraction of green valley galaxies that varies from $\approx 20$\% at $z=1$ to $\approx 10$\% at $z=0$. We remind the reader that D1 and D2 definitions of ${\tau}_{rm Q}$ diverge most noticeably at high stellar masses, where the populations are most difficult to distinguish.

{ We expect the exact fractions of passive and green-valley galaxies to change depending on our definition D1 and D2. We investigate how much these change due to variations in our parameters outlined in $\S$~\ref{colorselec} and $\S$~\ref{sfrselec} in Appendix~\ref{appendixA}. We include the results in Fig. \ref{fig:Fraction_Check}. Although the normalisation of these fractions change considerably with different population definitions, we find the shape of the distributions with stellar mass remains consistent. As such, we are confident our conclusions are robust and agnostic to the parameters we use. }

\section{Discussion and conclusions}\label{conclusions}

In this paper, we have explored the connection between quenching timescales and several galaxy properties in the \eagle\ simulations, with the aim of connecting to the different physical processes behind the quenching of star formation in galaxies. We do this using two different definitions (D1, and D2) of quenching timescale. Our study reveals some clear dependencies, which we summarise below. 

\begin{itemize}

\item Regarding the two different definitions of quenching timescale, we find that D1 leads to systematically longer quenching timescales than D2, as expected from the time delay between the ceasing of star formation and the reddening of passively evolving stellar populations. However, all of the conclusions below concerning key difference between centrals/satellites and environment/AGN feedback remarkably appear robust against both of these definitions.

\item Low-mass and high-mass central galaxies exhibit divergent quenching timescale distributions, where low mass centrals were most likely to quench with timescales ${\tau}_{\rm Q}\approx 3.5$~Gyr in D1, and there being a noticeable knee in the distribution at this ${\tau}_{\rm Q}$ value in D2. High mass centrals have a sharper peak at ${\tau}_{\rm Q}\lesssim 1$. 

\item Satellite galaxies have a noticeable knee in their  ${\tau}_{\rm Q}$ distribution at $1.5-3$~Gyr compared to high mass central galaxies ($<1.5$~Gyr) in \eagle. 

\item We find satellite and central galaxies in \eagle\ (that by $z=0$ are passive) have similar dependencies of their quenching timescales with their stellar mass, except in the lower stellar mass regime. Both populations reveal a quenching timescale that increases with stellar mass up to $M_{\star}\lesssim 10^{10}\,\rm M_{\odot}$, with the relation reversing at higher stellar masses. This said, at lower stellar mass values, central galaxies have a systematically longer quenching timescale than satellite galaxies, making the ${\tau}_{\rm Q}$ peak less dominant for centrals. {The amount of gas in the ISM of galaxies at the time a galaxy leaves either the blue cloud or the SFS is a strong predictor of quenching timescales, but more so for centrals than for satellites.}

\item For central galaxies, we find at lower stellar masses ($M_{\star}<10^{9.6}M_{\odot}$) that quenching timescales of galaxies are long compared to their satellite counterparts. We attribute the longer quenching timescales to be characteristic of the secular evolution of the stellar population, induced by stellar feedback. At intermediate stellar mass for centrals ($10^{9.7}M_{\odot}<M_{\star}<10^{10.3}M_{\odot}$), we find the peak in quenching timescales to be associated with galaxies which are gas-rich in their inner-most ($30$~kpc) regions, where we see ${\tau}_{\rm Q}>3.5$~Gyr. This result is expected as the gas can readily be converted into stars. In the higher stellar mass regime $M_{\star}>10^{10.3}M_{\odot}$, we observe SMBH accretion rates to increase rapidly along with the cumulative number of galaxy mergers, two likely interrelated factors. Quenching timescales shorten again in this stellar mass range to ${\tau}_{\rm Q}<1.5$~Gyr, where AGN-induced quenching appears to dominate.

\item For satellites in the low stellar mass regime ($M_{\star}<10^{9.6}\,\rm M_{\odot}$), we find that quenching timescales are particularly short, $<1$~Gyr, if their stellar mass is low compared to their host halo mass, while stellar-halo mass ratios in excess of $10^{-3.5}$ are linked with lengthened quenching timescales, up to $>2$~Gyr. We attribute the shortening of satellite quenching timescales relative to centrals in this regime to be induced by ram pressure stripping of satellite galaxies, stifling their star formation. 
More massive satellites begin to behave very similarly to massive central galaxies, and we find their AGN activity and mean major merger count to be significantly higher for $M_{\star}>10^{10.3}\,\rm M_{\odot}$ where quenching timescales shorten again in this stellar mass regime to ${\tau}_{\rm Q}<1.5$~Gyr.

\end{itemize}

With the discrete snapshots offered by the \eagle\ database, it was impossible to resolve sub-Gyr quenching timescales robustly using the method presented here, and as such, we could not compare the characteristic timescales of environmental and internal quenching mechanisms below that time cadence. However, we can state that AGN and environmental quenching shift the ${\tau}_{\rm Q}$ distribution to sub-Gyr scales, and in gas-rich galaxies without significant AGN activity, longer timescales of ${\tau_{\rm Q}}>4$~Gyr are common. These longer timescales are mostly associated with low mass central galaxies that quench by the action of stellar feedback alone. Both our work and that of \citet{Correa18} show that the colour transformation and quenching timescales can have a complex dependence on galaxy properties.

The predictions we present here will be subject to comparison to future observations. Very high spectral resolution spectroscopy, allowing the construction of accurate star formation histories, will be made possible with the James Webb Space Telescope (JWST - \citealt{Gardner06}; see \citealt{Kalirai18} for a recent discussion of the science objectives of JWST). Those histories will be key to confirm the quenching timescales reported here for the \eagle\ simulations. In addition, highly complete surveys focused on intermediate-redshift galaxies such as DEVILS \citep{Davies2018DEVILS} will also provide valuable insight into the properties of green-valley galaxies as a function of the group properties. Comparisons between gas-rich and gas-poor groups and the corresponding frequency of green valley galaxies would be useful in assessing the validity of our conclusions, which will be possible with the new generation of SKA \citep{Blyth2015} in  coming years.

\subsection*{Acknowledgements}

We thank Camila Correa for her valuable scientific input and sharing a submitted version of her manuscript. RW is funded by a Postgraduate Research Scholarship from UWA. CL has received funding from a Discovery Early Career Researcher Award (DE150100618) and by the ARC Centre of Excellence for All Sky Astrophysics in 3 Dimensions (ASTRO 3D), through project number CE170100013. We also thanks the Research Collaboration Awards at UWA in its 2018 program, which funded the visit of Dr. James Trayford to ICRAR. We acknowledge the Virgo Consortium for making their simulation data available. The \eagle\ simulations were performed using the DiRAC-2 facility at Durham, managed by the ICC, and the PRACE facility Curie based in France at TGCC, CEA, Bruyeres-le-Chatel.



\bibliographystyle{mnras}
\bibliography{mnras_template.bib} 


\appendix

\section{Predictions of the evolution of the fraction of passive and green valley galaxies}\label{appendixA}

\begin{figure*}
\centering
\includegraphics[width=0.58\textwidth]{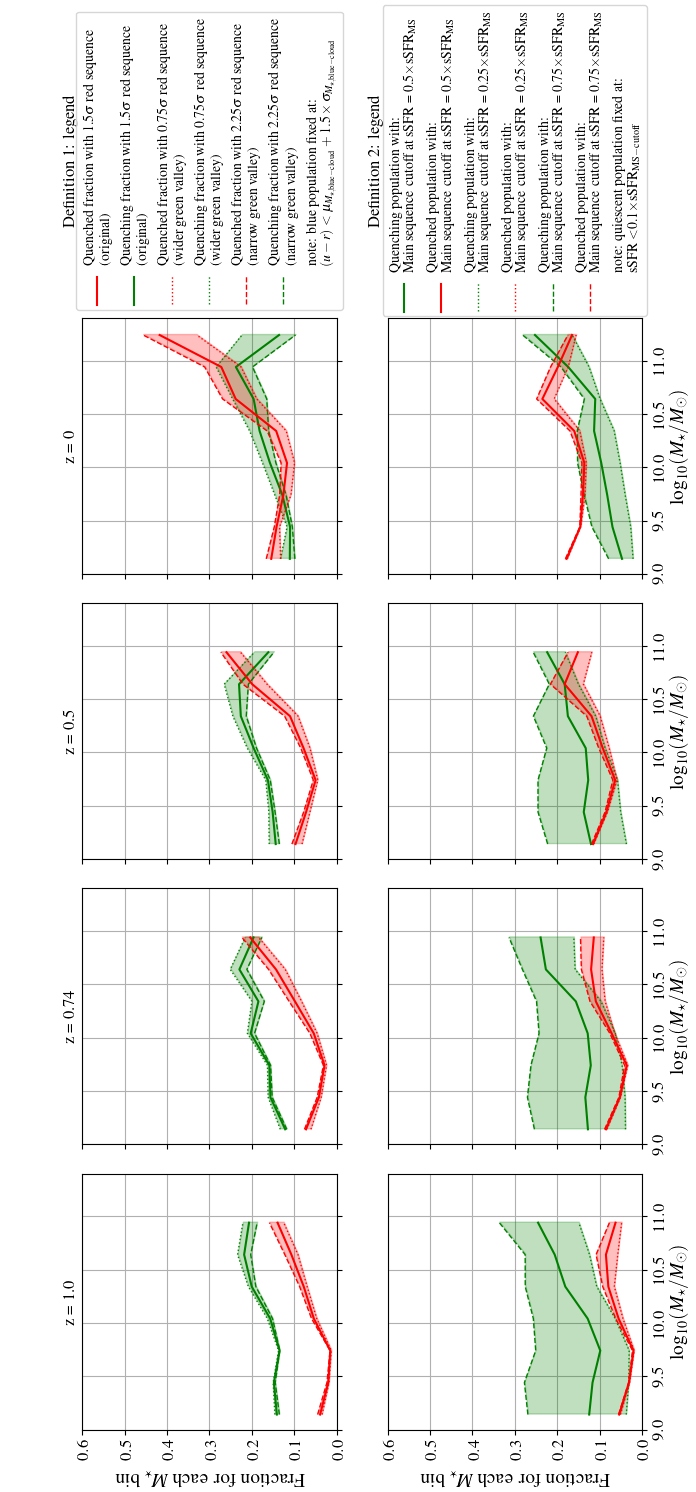}
\caption{The fraction of galaxies (as a function of stellar mass) that are part of the quenching or passive population in \eagle\ based on D1 (top 4 panels) and D2 (lower 4 panels) with altered versions of the definitions D1 and D2, for redshifts between $z=1$ and $z=0$. Shaded regions correspond to the range spanned by using different passive and star-forming galaxy definitions. Red lines and regions correspond to quenched fractions, while green lines and regions correspond to quenching/green valley fractions. Although the normalisation of these fractions change considerably with changing parameters, we find the shape of the distributions with stellar mass changes little, and hence is robust.}
\label{fig:Fraction_Check}
\end{figure*}

{Here, we test the robustness of the predicted fraction of passive and green valley galaxies and how those evolve in time. We do this by investigating how much these fractions change over stellar mass and redshift due to variations in the parameters outlined in $\S$~\ref{colorselec} and $\S$~\ref{sfrselec}. For D1, where we used a double-Gaussian to fit the colour bimodality, we  vary the $\sigma$ multiplier used to define the width of the red sequence to illustrate the effect this would have on quenched/quenching fractions. We changed the multiplier by $50$\% from $1.5$ to $0.75$ and $2.25$, the resulting fraction ranges shown in the top 4 panels of Fig.~\ref{fig:Fraction_Check}, where red and green represent the quenched and quenching fractions, respectively. Although the normalisation of these fractions changes slightly as a result of varying $\sigma$, we find that the shape of the quenched and quenching fractions over stellar mass remain consistent. 

For D2, we defined the star-forming and quiescent populations with appropriate multiplying factors ($0.5$ and $0.05$, respectively) of the star-forming main sequence, as outlined in $\S$~\ref{sfrselec}. We changed these multiplying factors by $50$\% to $0.75$ (and $0.075$) and $0.25$ (and $0.025$) to illustrate the effect this would have on our quenched and quenching fractions over stellar mass and redshift. We again find that the normalisation of these fractions can change (especially for the ``green'' population), however we see that the shape of said fractions remains consistent. Thus, we conclude that the predictions presented in $\S$~\ref{sec:predictions} are robust.}

\bsp	
\label{lastpage}
\end{document}